# Compressional rate-dependent stability of ammonia hydrates crystallized from water-rich ammonia-water solutions

A. Mondal[1,2 *], K. Mohrbach[1,3], K. Glazyrin[3], H.-P. Liermann[3], C. Sanchez-Valle[1]

[1] Institut für Mineralogie, Universität Münster, 48149 Münster, Germany

[2] Department of Physics, NIT Agarata, 799046 Jirania, India

[3] Deutsches Elektronen-Synchrotron DESY, Notkestr. 85, 22607 Hamburg, Germany

*** Author to whom correspondence should be addressed:** anshuman.phy@faculty.nita.ac.in

## Abstract

Understanding the crystallization pathways of water-rich ammonia-water ($NH_3$-$H_2O$) solutions and the stability of ammonia hydrates is key to unraveling the behavior of complex hydrogen-bonding networks as well as for planetary interior modelling. Yet, there are still inconsistencies in the crystallization sequence reported upon pressure-induced crystallization of $H_2O$-rich $NH_3$-$H_2O$ solutions at room temperature. Here, we investigate the effect of compression rates on the crystallization pathways of 25 wt% $NH_3$ aqueous solutions at room temperature using dynamically compressed diamond anvil cells (dDAC) coupled with time-resolved X-ray diffraction. We show that compression rates exceeding 0.5 GPa/sec promote direct crystallization of a body-centered cubic (bcc) phase (DMA') with possible AMH stoichiometry coexisting with $H_2O$ ice VII, while rates below 0.2 GPa/sec stabilize monoclinic $NH_3$-rich AHH-II and ice VII phases. Intermediate rates between 0.2-0.5 GPa/sec produce a mixture of both hydrates alongside ice VII, hence demonstrating the role of compression rate on the crystallization sequence of ammonia solutions. The compression behavior and phase stability of the distinct phase assemblies (AHH-II/DMA' + ice VII) are investigated further to place constraints into the composition of the DMA' phase, the effect of ice VII on the compressibility of ammonia hydrates and the plausible incorporation of $NH_3$ impurities within the lattice of high-pressure ice phases.

# I. Introduction

Understanding the phase behavior of the ammonia-water ($NH_3$-$H_2O$) system under extreme pressure-temperature (P-T) conditions is of critical importance in both materials physics[1] and planetary science[2]. As a model hydrogen-bonded molecular system[1], these binary mixtures exhibit crystallization behavior that is sensitive to stoichiometry and thermal history, making them ideal for studying metastable phase formation and kinetic effects under compression[3,4]. Furthermore, as important constituents of the mantle icy moons[5] (e.g. Titan, Europa, etc.), the ice giant planets[6–8] (Uranus and Neptune) and ice-rich exoplanets[9–11], the stability of phases in the $NH_3$-$H_2O$ system over a broad range of pressure-temperature conditions is critical in modeling their internal structure and thermal evolution. At room pressure, the crystallization of $NH_3$-$H_2O$ solutions stabilizes three stoichiometrically distinct ammonia hydrate forms[12]: ammonia hemihydrate (AHH, $2NH_3$:$1H_2O$), ammonia monohydrate (AMH, $1NH_3$:$1H_2O$) and ammonia dihydrate (ADH, $1NH_3$:$2H_2O$). Despite extensive research efforts, significant discrepancies persist within the reported P-T phase diagrams[1,13–16], particularly in relation to the nucleation behavior[12,13,16] and phase stability[14–18] of ammonia hydrates. Upon compression at room temperature, $NH_3$-$H_2O$ solutions undergo fractional crystallization that first stabilizes $H_2O$ ice VII, followed by the crystallization of the residual $NH_3$-rich liquid into an ammonia hydrate phase whose nature remains a matter of debate[14,15]. For instance, the formation of the monoclinic AHH-II phase has been identified at pressure above 3.5 GPa[17,19,20], while studies by Boone et al.[15,16] have reported the formation of an AMH phase and $H_2O$ ice VII upon crystallization of $NH_3$-$H_2O$ solutions. Moreover, Wilson et al.[14] have reported the direct formation of an unidentified body-centered cubic (bcc) disordered molecular alloy (DMA) phase, unlike the AHH-II, in coexistence with ice VII at pressures as low as 7.6 GPa under rapid compression from 32 wt% $NH_3$-$H_2O$ solutions at room temperature. These observations suggest that compression rates may significantly influence the phase relations, nucleation kinetics, and the crystallization pathways of novel stable or metastable phases[21,22]. Despite this potential importance, to the best of our knowledge, the role of compression rate in the crystallization of the $NH_3$-$H_2O$ solutions has yet to be systematically explored.

To address this gap, we utilized diamond anvil cells with dynamic compression capabilities[23], in combination with time-resolved X-ray diffraction (XRD), to investigate the effect of compression rates ranging from 0.005 to 4 GPa/sec on the crystallization of 25 wt% $NH_3$-$H_2O$ solutions at room temperature, 298 K. Our results reveal the formation of three distinct phase assemblages depending upon compression rate: 1) monoclinic AHH-II and ice

VII assembly below 0.2 GPa/sec; 2) a bcc DMA phase (hereafter DMA') and ice VII assembly above 0.5 GPa/sec and 3) AHH-II + DMA' + ice VII between 0.2 and 0.5 GPa/sec. The compression responses of the AHH-II + ice VII and DMA' + ice VII phase assemblages were examined further up to 93 GPa and 146 GPa, respectively. These findings prompt further discussion into the structural identity of the DMA' phase, the influence of ice VII on the equation of state (EoS) of ammonia hydrates, and the potential incorporation of $NH_3$ impurities within the bcc $H_2O$ ice VII lattice.

# II. Experimental methods

## A. Diamond anvil cell (DAC) preparations

Symmetric diamond anvil cells (DACs) equipped with type Ia diamonds with 150-200 μm culet diameter were utilized for the experiments. Rhenium (Re) gaskets, which were pre-indented to an approximate thickness of 30 μm and drilled with a central hole of half the size of diamond culet diameter, were employed as sample chambers. Compression runs were carried out using commercially available $NH_3$-$H_2O$ solutions containing 25 wt% $NH_3$ (Sigma-Aldrich, product no. 1.05428). To prevent $NH_3$ evaporation from the solution during loading, the samples were refrigerated to approximately 278 K and the DAC pre-cooled to around 292 K before loading. A droplet of the solution was carefully placed into the sample chamber with a medical syringe, and the DAC was immediately closed and pressurized. Fine-grained gold (Au) powders were introduced into the compression chamber to serve as pressure markers during the compression experiments.

Two different sets of experiments were performed in this study, namely the compression rate-dependent crystallization experiments on 25 wt% $NH_3$-$H_2O$ solutions and thereafter the P-V compression experiments on the identified crystalline assemblages. For the crystallization experiments, after placing the solution into the compression chamber, the DACs were immediately closed at pressures below 1.5 GPa so that the entire sample remained in the liquid state (**Table I**). Thereafter, a piezo-driven dynamic DAC (dDAC)[23] was used to accurately control the compression rate during the crystallization experiments. A total of eleven dDAC runs were performed by applying trapezoidal voltage waveform with compression rates ranging from 0.005 to 4 GPa/sec (**Table I**). For the P-V compression experiments, 25 wt% $NH_3$-$H_2O$ solutions were loaded into the compression chambers, and the DACs were mechanically pressurized with undefined compression rates in order to stabilize the distinct crystallized assemblies identified in the compression rate-dependent crystallization studies. The sample in run dDAC 13 (**Table II**), was partially crystallized, i.e. with formation of $H_2O$ ice VII only, before the experiments. Further compression was achieved through either a gas-driven membrane (mDAC) or using the dDAC[23] setup described above. For mDAC runs, pressure was increased in steps of ~0.5 GPa corresponding to the membrane gas pressure increase of ~0.2 bar and with a slew rate of 0.05 bar/sec (**Table II**). In the dDAC runs, a triangular voltage waveform was selected for ramp compressions with compression rates listed in **Table II**.

## B. X-ray diffraction data collection and analysis

Angle dispersive powder x-ray diffraction experiments were conducted at beamline P02.2 of PETRA III, DESY, Hamburg, Germany[24]. Powder XRD patterns were collected during compression runs using monochromatic synchrotron x-rays focused to a beam spot size of 3 x 8 $\mu m^2$ full-width at half maxima (FWHM). The compression rate-dependent crystallization experiments were conducted employing an x-ray energy of 25.63 keV (**Table I**) while the incident x-ray energies for the four P-V compression experiments are listed in **Table II**. In the crystallization experiments, time-resolved XRD patterns were collected utilizing a Perkin Elmer XRD1621 flat panel detector with an exposure time of 1 sec at different compression rates. Moreover, 2-dimensional (2D) XRD maps were recorded across the entire compression chamber with an exposure time of 0.2 sec at the maximal pressure of each crystallization runs to identify the synthesized crystalline assemblages (**Table I**). For the P-V compression experiments, the XRD data were collected using the Perkin Elmer XRD1621 flat panel detector and two 2.3 MPix GaAs LAMBDA detectors[25,26] for mDAC and dDAC runs, respectively. The data acquisition times for all the four compression runs are listed in **Table II**. Moreover, before compression, XRD maps were collected across the entire compression chamber with an exposure time of 0.5-1 sec.

The calibration of sample-to-detector distance, tilt and rotation for XRD1621 or LAMBDA detectors were accomplished using x-ray standards ($CeO_2$ or $Cr_2O_3$ from NIST SRM 674b). The 2D diffractograms were integrated into 1-dimensional (1D) diffraction patterns using the DIOPTAS software[27]. A quick visualization through an intensity plot (**Figure 1**) stacking all diffractograms collected during the compression run was generated using the modified DIOPTAS software integrated in the P02.2 data pipeline[28]. The 2θ-angle of individual reflection was determined based on a pseudo-Voigt fit using a python least-squares fitting routine[29] (https://gitlab.desy.de/rachel.husband/peakfit).

In the P-V compression experiments (**Table II**), the weak diffraction signal of the low-Z ammonia hydrate samples and the limited angular coverage offered by the 2.3 MPix LAMBDA detectors resulted in reduced 2θ-angle sampling in the dDAC runs compared to the mDAC experiments. Therefore, only the (110) reflection of the bcc ammonia hydrate phases was observed throughout the entire pressure range in the dDAC experiments, compared to the two to three reflections in the static mDAC runs (**Figures S1** and **S2**). To facilitate direct comparison among all four datasets (**Table II**), the lattice parameters of the bcc AHH-D(I)MA and DMA' phases were determined using only the position of the (110) reflection. The uncertainty in volume of the bcc ammonia hydrate phases was estimated based on the standard deviation in the volumes determined from (110), (211) and/or (220) reflections in the diffraction data collected in the mDAC runs in the low-pressure ranges (**Figure S1**, **S2**, **Table II**). We assumed the error in the volumes to be similar for the dDAC compression runs, as well as for the high pressure mDAC XRD data. Pressure determination relied on the position of the Au (111) reflection[30] that is known to be least sensitive to nonhydrostatic uniaxial stress[31]. The bulk modulus ($B_0$) and its pressure-derivative ($B_0'$) at ambient temperature were derived from fitting the dense P-V datasets to a 3rd Birch-Murnaghan equation of state (BM EoS) using the EoSFIT7C program[32].

# III. Results and Discussion

## A. Compressional rate-dependent stability of ammonia hydrates

The outcomes of the compression-rate dependent crystallization experiments of the 25 wt% $NH_3$-$H_2O$ solutions (dDAC runs) are summarized in **Table I**. The XRD maps recorded at the highest pressure plateau (**Figure S3**), along with the intensity plots displaying the stacking of integrated diffractograms collected during the ramp compression runs (**Figure 1**) reveal the formation of a bcc DMA-like phase[14], hereafter referred to as DMA', coexisting with ice VII, at compression rates exceeding 0.5 GPa/sec. This ammonia hydrate phase is identified by the characteristic bcc (110) reflection present at the 2θ angle of ~11.79° (**Figure 2**). Although neutron diffraction data is unavailable to confirm the disordered nature of the phase, we infer this phase to be DMA-like, which appears to be thermodynamically stable phase of ammonia hydrates at high pressure[14]. Additionally, the FWHM of the DMA' reflections are found to be higher in comparison to the reflections arising from known ordered structures (AHH-II or Au) in the XRD spectra, further supporting the hypothesis of intrinsic disorder. In contrast, compression rates below 0.2 GPa/sec stabilize the monoclinic AHH-II structure[14,17], in coexistence with $H_2O$ ice VII, as indicated by the characteristics $(12\bar{1})$, $(102)$ and $(023)$ reflections in between 11 to 12 deg $2\theta$ angle region (**Figure 2**). Interestingly, compression rates between 0.2 and 0.5 GPa/sec yield a mixture of both ammonia hydrate phases (DMA' + AHH-II), together with $H_2O$ ice VII, which are inhomogeneously distributed within the compression chamber (**Figure S3**). However, the fractions of the individual ammonia hydrate phases could not be extracted from the recorded diffractograms as the presence of large crystallites prevented the formation of homogeneous powder rings. Representative XRD patterns recorded at compression rates of 0.005, 0.2 and 2 GPa/sec, together with their corresponding Le Bail refinements are displayed in **Figure 2**.

These observations confirm that compression rate plays a pivotal role in the crystallization pathway of ammonia-water solutions containing 25 wt% $NH_3$, leading to the formation of distinct phase assemblages. The present results thus resolve discrepancies in the phase relations previously reported for the crystallization of $H_2O$-rich $NH_3$-$H_2O$ solutions[14]. An updated pressure-temperature phase diagram for the 25 wt% $NH_3$-$H_2O$ system, modified after Mousis et al[13] and Wilson et al[14], is presented in **Figure 3**. This revised phase diagram provides a more comprehensive understanding of the compression-dependent stability of ammonia hydrates crystallized from $H_2O$-rich $NH_3$-$H_2O$ solutions and serves as a foundation for further investigation into their structural and thermodynamic properties. While the stoichiometry of the DMA' phase cannot be directly inferred in the absence of single-crystal XRD data, we present an estimate of the composition by investigating its compressional behavior in comparison with the P-V curves of other known DMA/DIMA phases in the $NH_3$-$H_2O$ system, as presented below.

### B. Pressure evolution of the AHH-II and DMA' phases

Selected diffractograms recorded upon compression of the AHH-II + ice VII and DMA' + ice VII assemblies in both membrane-driven mDAC (mDAC1, mDAC2) and piezo-driven dDAC runs (dDAC12, dDAC13) are presented in **Figure S1** and **S2,** respectively. In mDAC1 and dDAC12 runs (**Figure S1**), the crystallized ammonia hydrate phase is identified as monoclinic AHH-II[17]. Upon further compression above 19 GPa, these three reflections merge into a single broad reflection, indicating the transformation into the bcc AHH-D(I)MA structure (**Figure S1**). The transition pressure aligns well with previous reports[14,19,33,34] despite the presence of ice VII in the sample chamber. The phase appears stable up to 93 GPa without detectable changes in the pattern that would indicate the spontaneous partial ionization of the phase reported above 26 GPa on the basis of Raman spectroscopic measurements[34,35]. Because XRD analysis is not sufficiently sensitive to changes in the hydrogen bonding network, the partially ionic characteristics of the AHH-D(I)MA phase could not be resolved. We note that to date, evidence of partial ionization has only been detected in the ammonia monohydrate (AMH)-DIMA phase using XRD measurements. Moreover, the transformation to a purely ionic $P\bar{3}m1$ structure above 69 GPa[35] was not observed either.

The XRD spectra collected in run mDAC2 confirmed the crystallization of the bcc-structured ammonia hydrate DMA' and ice VII phases upon initial rapid mechanical compression of the 25wt% $NH_3$-$H_2O$ solution (**Figure S2a**). Whereas, for run dDAC13 initial compression up to 2.5 GPa resulted in partial crystallization of the $NH_3$-$H_2O$ solution (**Figure S2b**). Only upon further compression above ~4 GPa, the DMA' phase and ice VII assembly became stable. Analysis of both mDAC2 and dDAC13 runs demonstrate that the bcc structure of DMA' phase is stable up to 146 GPa, the highest pressure investigated in these experiments (**Table II**). Similar to our previous discussion, Raman spectroscopic measurements will be necessary to confirm whether partial ionization of this phase occurred at higher pressures. As reported in other studies on DMA/DIMA phases of the $NH_3$-$H_2O$ system[36,37], their partially ionic counterparts, if present, may exist in minor fractions within the compression chamber and could remain undetected in XRD patterns due to their very weak effect to the overall scattering signal.

**Table I.** Summary of experimental conditions, compression rates and characterized phase assemblies observed in the crystallization experiments of 25 wt% $NH_3$-$H_2O$ solutions.

| Run No | Compression rate (GPa / sec) | Starting P (GPa) | Maximal P (GPa) | Phases |
|---|---|---|---|---|
| **dDAC1** | 4 | 0.45 | 13 | DMA' + ice VII |
| **dDAC2** | 2 | 0.3 | 9.4 | DMA'+ ice VII |
| **dDAC3** | 0.5 | 1 | 6.6 | AHH-II + DMA'+ ice VII |
| **dDAC4** | 0.5 | 1.2 | 10.4 | AHH-II + DMA'+ ice VII |
| **dDAC5** | 0.2 | 0.6 | 19.2 | AHH-II + DMA'+ ice VII |
| **dDAC6** | 0.2 | 1.3 | 9.4 | AHH-II + DMA'+ ice VII |
| **dDAC7** | 0.2 | 0.4 | 11.6 | AHH-II + DMA'+ ice VII |
| **dDAC8** | 0.1 | 0.2 | 5.6 | AHH-II+ ice VII |
| **dDAC9** | 0.05 | 1.2 | 3.8 | AHH-II+ ice VII |
| **dDAC10** | 0.01 | 0.9 | 8.2 | AHH-II+ ice VII |
| **dDAC11** | 0.005 | 0.4 | 13.2 | AHH-II+ ice VII |

**Table II:** Summary of P-V compression runs performed on AHH-II and DMA' phases

| Run No | X-ray energy (keV) | Exposure time (sec) | Starting P (GPa) | Maximal P (GPa) | Slew (mDAC)/ Compression (dDAC) rate | Starting assemblages |
|---|---|---|---|---|---|---|
| **mDAC1** | 25.62 | 5 | 5.8 | 71.1 | 0.05 bar/sec | AHH-II + ice VII |
| **mDAC2** | 42.65 | 5-10 | 8.8 | 146.4 | 0.05 bar/sec | DMA'+ ice VII |
| **dDAC12** | 25.59 | 0.2 | 6 | 92.8 | 1 GPa/sec | AHH-II + ice VII |
| **dDAC13** | 25.62 | 2 | 2.5 | 36.6 | 0.1 GPa/sec | DMA'+ ice VII |

### C. Plausible stoichiometry of DMA' based on P-V compression curves

The pressure dependence of the DMA' phase volumes (per molecule) obtained from both mDAC and dDAC compression runs (**Table II, Table S1**) are depicted in **Figure 4** together with those for AHH-D(I)MA. The P-V data of the AHH-D(I)MA phase are in good agreement with previous literature data[14,33,34]. Moreover, the volume (per molecule) of AHH-D(I)MA is found to be higher than that of the bcc DMA' phase, with differences of approximately 7% at 30 GPa (**Figure 4**). In order to identify the probable stoichiometric composition of the DMA' phase, we compared its volume with all known stable ammonia hydrate DMA and/or DIMA phases[14,29,33,37–39] as well as with the end members $H_2O$[40] and $NH_3$[41] (**Figure 4**). Because the compression curve of the DMA' phase lies closer to the pure ice VII curve than that of the AHH-D(I)MA phase, it indicates a higher $H_2O$ content in the DMA' phase. The composition of the DMA' phase can be further assessed tentatively by comparing its compression curve with that of the AMH-DIMA[38] phase, which has a known 1:1 $NH_3$ to $H_2O$ molar ratio. The good agreement between these two compression curves within their respective uncertainties, especially at pressures below 30 GPa and above 60 GPa **(Figure 4)**, suggest a $1NH_3$:$1H_2O$ composition for the DMA' phase. The differences in the intermediate pressure range of the P-V curves may be due to the presence of $H_2O$ ice VII in the compression chamber, which may develop locally distinct stress conditions[42,43] that affect the elastic response of the DMA' phase. Alternatively, this difference may also be interpreted as a possible loss of $NH_3$ from the DMA' crystal lattice incorporating it into the coexisting $H_2O$ ice VII, which will be discussed in the following section.

In order to conclusively identify the origin of such deviation in the volume compression curves, at least a line shift analysis[44] would have to be performed in order to assess the stress of the phases and possible non-hydrostaticity, which with the existing data quality is not possible. Nevertheless, it is worth noting that our DMA' phase volumes are in excellent agreement with the P-V data reported by Wilson et al.[14] for the bcc DMA phase observed in coexistence of $H_2O$ ice VII crystallized upon rapid compression of 32 wt% $NH_3$-$H_2O$ solutions (**Figure 4**). Despite the lack of single-crystal structural refinements and/or a chemical analysis, the strong similarities in the compressional behavior of the DMA' and AMH-DIMA phases suggest that they can be treated as equivalent. However, signatures of the partially ionic counterpart of the DMA'/AMH-DIMA phase could not be identified from our XRD data as opposed to previous reports[36,45]. This suggests a distinct phase relation of DMA phases in presence of excess $H_2O$ ice VII.

Finally, we compared the P-V data for the DMA' phase with that of an ideal AMH-DIMA phase (hereafter referred to as "AMH ideal") calculated by linear mixing of the $H_2O$ ice VII[40] and $NH_3$[41] volumes **(Figure 4)**. It is worth noting that the compression behavior of the DMA'/AMH-DIMA phase closely follows the ideal curve, particularly at low pressures (<20 GPa), which contrasts with the results from a previous compression study on the ADH-DMA phase[29]. This discrepancy in compression behavior at low pressures between the two different compositions could potentially originate from the complexity of the H-bonding network[39] and might be linked to the higher $H_2O$ concentration in the ADH-DMA phase, which causes it to deviate from its ideal behavior.

### D. Equation of state of DMA/DIMA phases in the $NH_3$-$H_2O$ system

In **Figure 5**, we fitted the P-V datasets for the AHH-D(I)MA and DMA' phases **(Table S1)**, including their associated errors, to a 3rd-order BM EoS for comparison with previous studies. We adopted initial values of the ambient pressure volume ($V_0$) assuming an ideal mixing between $NH_3$[41] and $H_2O$[40] with a 2:1 molar ratio for AHH-D(I)MA and a 1:1 ratio for DMA', and later refined the $V_0$ values until convergence. The best-fit EoS parameters are compared with all three stable ammonia hydrate DMA/DIMA forms in **Table III**. The fit parameters obtained for the AHH-D(I)MA phase are in good agreement with previously reported values[34]. For both AHH-D(I)MA and DMA' systems, the entire P-V dataset could be described with a single EoS-fit. The EoS parameters for the DMA' phase display a higher bulk modulus value compared to both the AHH-D(I)MA and ADH-DMA phases, but are in close agreement to those of AMH-DIMA phase[38] **(Table III)**, further supporting the similar nature and composition of the DMA' and AMH-DIMA phase.

**Table III.** Birch-Murnaghan Equation of state parameters for ammonia hydrate DMA/DIMA phases in the $NH_3$-$H_2O$ system.

| EoS parameters | AHH-D(I)MA (this work) | AHH-DIMA[34] | DMA' (this work) | AMH-DIMA[38] | ADH-DMA[29] |
|---|---|---|---|---|---|
| $V_0$ (Å$^3$/molecule) | 29.25 ± 0.07 | 29.1 ± 0.2 | 26.06 ± 0.03 | 25.93 ± 0.03 | 23.96 ± 0.03 |
| $B_0$ (GPa) | 9.89 ± 0.01 | 9.6 ± 0.3 | 11.38 ± 0.05 | 12.17 ± 0.05 | 9.95 ± 0.14 |
| $B'_0$ | 4.85 ± 0.02 | 4.6 ± 0.3 | 5.3 (fixed) | 5.3 (fixed) | 6.59± 0.03 |

## E. Incorporation of $NH_3$ impurities in $H_2O$ ice and effects on its compressibility

The present work also permits to examine the high-pressure behavior of $H_2O$ ice in the presence of ammonia hydrates, and particularly, whether $NH_3$ molecules may be incorporated into the crystal lattice of ice VII. The P-V data obtained for $H_2O$ ice in all four compression runs (**Table II, Table S2**) are compared in **Figure 6**. The results from mDAC1 and dDAC12 runs in presence of the AHH-II/AHH-D(I)MA phase (**Table II**) are in good agreement with the pure $H_2O$ ice data previously reported in the literature[40,46]. This agreement is expected because the more compressible AHH[34], with its lower bulk modulus, would not significantly influence the compression behavior of $H_2O$ ice[40] crystals. However, the $H_2O$ ice volume data collected in mDAC2 and dDAC13 runs, where it coexists with the DMA' phase (**Table II**), deviates from the other datasets between 30 and 60 GPa, with maximal differences of ca. 3% at 38 GPa. This observation thus suggests a different compression behavior of $H_2O$ ice depending on whether it coexists with the AHH-D(I)MA or the DMA' phase in the sample chamber. Since DMA' is less compressible than AHH and has a compressibility somewhat closer to that of $H_2O$ ice VII, its presence likely alters the overall compression behavior of $H_2O$ ice observed in these runs.

While we cannot rule out that the observations are related to deviatoric stresses developed in the compression chamber by the mechanical contact between $H_2O$ ice and DMA' phases, the deviations could be also explained by the incorporation of $NH_3$ impurities into the ice VII crystal lattice, consistent with the deviation observed in the compression curve of the DMA' phase. A possible reason for this is the higher $H_2O$ content in the DMA' crystal structure compared to the AHH phase, which could result in a less densely packed hydrate structure[14,17]. This behavior originates from weaker interactions between $NH_3$ and the $H_2O$ ice frameworks, promoting $NH_3$ loss and its subsequent trapping in the $H_2O$ ice lattice. Although there are no previous reports on $NH_3$ incorporation into ice VII at high pressures, earlier XRD experiments have suggested the possibility of incorporating up to 1.6 mol% NaCl, 5 mol% $CH_3OH$ and 1.8 mol% $CaCl_2$ into the interstitial sites of the bcc $H_2O$ ice VII crystal lattice, which resulted in volume changes that are comparable to this study, i.e. -4.5 % for NaCl[47,48], +1.2% for $CH_3OH$[49] and -3% for $CaCl_2$[43]. This variation in volume change arises from the distinct chemical pressures exerted by the incorporated species[43]. While larger cations like $Ca^{2+}$ and $Na^{+}$ induce lattice compression, i.e. volume reduction, $CH_3OH$ incorporation leads to expansion of the ice VII lattice. Similarly, if $NH_3$ was incorporated, its larger radius would decrease the volume by exerting stronger chemical pressure on the lattice, potentially leading to compression. Overall, its strong hydrogen bonding interactions could also influence the lattice structure. Above 60 GPa, the $H_2O$ ice volume is in better agreement within all four different compression runs (**Figure 6**), and consistent with data reported for $H_2O$ ice X' and ice X[46], suggesting a more ideal behavior of the $NH_3$ impurities in the $H_2O$ ice lattice if the incorporation occurred or a loss of $NH_3$.

# IV. Conclusion

Our study provides a comprehensive understanding on the influence of compression rates on the crystallization pathways of $H_2O$-rich $NH_3$-$H_2O$ solutions (25 wt% $NH_3$), resolving discrepancies in the previously reported phase relations and crystallization processes. With the Le Bail refinements, we demonstrate that compression rates significantly affect the crystallization of distinct ammonia hydrate phases, with rates above 0.5 GPa/sec favoring the direct formation of the bcc DMA' phase, while rates below 0.2 GPa/sec stabilize the monoclinic AHH-II phase, both coexisting with $H_2O$ ice VII. We show that AHH-II transforms into the bcc AHH-D(I)MA phase above 19 GPa, in agreement with previous studies, and thereafter this phase remains stable up to 93 GPa in the presence of $H_2O$ ice VII. In contrast, the DMA' phase was identified to remain stable up to 146 GPa without any evidence of phase transformations. Our detailed P-V comparison of these two assemblages suggested that the DMA' phase displays higher $H_2O$ content compared to AHH. Based on the pressure-evolution of the DMA' phase volume, we conclude that this phase is closely related to the AMH-DIMA phase, despite the lack of confirmation via single-crystal diffraction or chemical analysis. Additionally, our results suggest the incorporation of $NH_3$ into the $H_2O$ ice VII lattice at high pressures, contributing to a deeper understanding of the stability and structural transitions of ammonia hydrates under extreme conditions. The implications of these findings extend beyond the fundamental behavior of hydrogen bonded system under compression, providing valuable insights into the role of compression rates in crystallization and metastable phase stability in $NH_3$-$H_2O$ mixtures. These insights are crucial for modeling the phase behavior of $H_2O$-rich planetary bodies and evidence the complex chemistry and physics of icy exoplanets and moons.

# Acknowledgments

This research was supported by the German Science Foundation (Deutsche Forschungsgemeinschaft, DFG) through Research Unit FOR 2440/2 (Grant No. SA2585/5-1). We acknowledge DESY (Hamburg, Germany), a member of the Helmholtz Association, for providing access to the experimental facility PETRA III and beamline P02.2. We also acknowledge the scientific exchange and support of the Centre for Molecular Water Science (CMWS). We also want to thank P02.2 beamline engineer M. Wendt for his useful technical support.

## Data and code availability

The collected raw XRD data is available upon request to the corresponding author. The python routine employed to extract the peak position of individual reflection from the XRD patterns is available at: https://gitlab.desy.de/rachel.husband/peakfit.

## References

[1] J.S. Loveday, and R.J. Nelmes, "The ammonia hydrates—model mixed-hydrogen-bonded systems," High Pressure Research **24**(1), 45–55 (2004).

[2] A.D. Fortes, and M. Choukroun, "Phase Behaviour of Ices and Hydrates," Space Sci Rev **153**(1–4), 185–218 (2010).

[3] C.M. Pépin, R. André, F. Occelli, F. Dembele, A. Mozzanica, V. Hinger, M. Levantino, and P. Loubeyre, "Metastable water at several compression rates and its freezing kinetics into ice VII," Nat Commun **15**(1), 8239 (2024).

[4] Y.-H. Lee, J.K. Kim, Y.-J. Kim, M. Kim, Y.C. Cho, R.J. Husband, C. Strohm, E. Ehrenreich-Petersen, K. Glazyrin, T. Laurus, H. Graafsma, R.P.C. Bauer, F. Lehmkühler, K. Appel, Z. Konôpková, M. Tang, A.P. Dwivedi, J. Sztuck-Dambietz, L. Randolph, K. Buakor, O. Humphries, C. Baehtz, T. Eklund, L.K. Mohrbach, A. Mondal, H. Marquardt, E.F. O'Bannon, K. Amann-Winkel, C.-S. Yoo, U. Zastrau, H.-P. Liermann, H. Nada, and G.W. Lee, "Multiple freezing–melting pathways of high-density ice through ice XXI phase at room temperature," Nat. Mater., (2025).

[5] H. Hussmann, C. Sotin, and J.I. Lunine, "Interiors and Evolution of Icy Satellites," in *Treatise on Geophysics*, (Elsevier, 2015), pp. 605–635.

[6] W.B. Hubbard, and J.J. MacFarlane, "Structure and evolution of Uranus and Neptune," J. Geophys. Res. **85**(B1), 225–234 (1980).

[7] J.J. Fortney, and N. Nettelmann, "The Interior Structure, Composition, and Evolution of Giant Planets," Space Sci. Rev. **152**(1–4), 423–447 (2010).

[8] R. Helled, N. Nettelmann, and T. Guillot, "Uranus and Neptune: Origin, Evolution and Internal Structure," Space Sci. Rev. **216**(3), 38 (2020).

[9] W.J. Borucki, D. Koch, G. Basri, N. Batalha, T. Brown, D. Caldwell, J. Caldwell, J. Christensen-Dalsgaard, W.D. Cochran, E. DeVore, E.W. Dunham, A.K. Dupree, T.N. Gautier,

J.C. Geary, R. Gilliland, A. Gould, S.B. Howell, J.M. Jenkins, Y. Kondo, D.W. Latham, G.W. Marcy, S. Meibom, H. Kjeldsen, J.J. Lissauer, D.G. Monet, D. Morrison, D. Sasselov, J. Tarter, A. Boss, D. Brownlee, T. Owen, D. Buzasi, D. Charbonneau, L. Doyle, J. Fortney, E.B. Ford, M.J. Holman, S. Seager, J.H. Steffen, W.F. Welsh, J. Rowe, H. Anderson, L. Buchhave, D. Ciardi, L. Walkowicz, W. Sherry, E. Horch, H. Isaacson, M.E. Everett, D. Fischer, G. Torres, J.A. Johnson, M. Endl, P. MacQueen, S.T. Bryson, J. Dotson, M. Haas, J. Kolodziejczak, J. Van Cleve, H. Chandrasekaran, J.D. Twicken, E.V. Quintana, B.D. Clarke, C. Allen, J. Li, H. Wu, P. Tenenbaum, E. Verner, F. Bruhweiler, J. Barnes, and A. Prsa, "Kepler Planet-Detection Mission: Introduction and First Results," Science **327**(5968), 977–980 (2010).

[10] N.M. Batalha, J.F. Rowe, S.T. Bryson, T. Barclay, C.J. Burke, D.A. Caldwell, J.L. Christiansen, F. Mullally, S.E. Thompson, T.M. Brown, A.K. Dupree, D.C. Fabrycky, E.B. Ford, J.J. Fortney, R.L. Gilliland, H. Isaacson, D.W. Latham, G.W. Marcy, S.N. Quinn, D. Ragozzine, A. Shporer, W.J. Borucki, D.R. Ciardi, T.N. Gautier, M.R. Haas, J.M. Jenkins, D.G. Koch, J.J. Lissauer, W. Rapin, G.S. Basri, A.P. Boss, L.A. Buchhave, J.A. Carter, D. Charbonneau, J. Christensen-Dalsgaard, B.D. Clarke, W.D. Cochran, B.-O. Demory, J.-M. Desert, E. Devore, L.R. Doyle, G.A. Esquerdo, M. Everett, F. Fressin, J.C. Geary, F.R. Girouard, A. Gould, J.R. Hall, M.J. Holman, A.W. Howard, S.B. Howell, K.A. Ibrahim, K. Kinemuchi, H. Kjeldsen, T.C. Klaus, J. Li, P.W. Lucas, S. Meibom, R.L. Morris, A. Prša, E. Quintana, D.T. Sanderfer, D. Sasselov, S.E. Seader, J.C. Smith, J.H. Steffen, M. Still, M.C. Stumpe, J.C. Tarter, P. Tenenbaum, G. Torres, J.D. Twicken, K. Uddin, J. Van Cleve, L. Walkowicz, and W.F. Welsh, "PLANETARY CANDIDATES OBSERVED BY *KEPLER* . III. ANALYSIS OF THE FIRST 16 MONTHS OF DATA," ApJS **204**(2), 24 (2013).

[11] L. Noack, I. Snellen, and H. Rauer, "Water in Extrasolar Planets and Implications for Habitability," Space Sci. Rev. **212**(1–2), 877–898 (2017).

[12] J.S. Kargel, "Ammonia-water volcanism on icy satellites: Phase relations at 1 atmosphere," Icarus **100**(2), 556–574 (1992).

[13] O. Mousis, J. Pargamin, O. Grasset, and C. Sotin, "Experiments in the $NH_3$ -$H_2$ O system in the [0, 1 GPa] pressure range - implications for the deep liquid layer of large icy satellites," Geophys. Res. Lett. **29**(24), (2002).

[14] C.W. Wilson, C.L. Bull, G.W. Stinton, D.M. Amos, M.-E. Donnelly, and J.S. Loveday, "On the stability of the disordered molecular alloy phase of ammonia hemihydrate," J. Chem. Phys. **142**(9), 094707 (2015).

[15] S. Boone, and M.F. Nicol, "Ammonia-water mixtures at high pressures - Melting curves of ammonia dihydrate and ammonia monohydrate and a revised high-pressure phase diagram for the water-rich region," Lunar and Planetary Science Conference Proceedings **21**, 603–610 (1991).

[16] M.L. Johnson, and M. Nicol, "The ammonia-water phase diagram and its implications for icy satellites," J. Geophys. Res. **92**(B7), 6339–6349 (1987).

[17] C.W. Wilson, C.L. Bull, G. Stinton, and J.S. Loveday, "Pressure-induced dehydration and the structure of ammonia hemihydrate-II," J. Chem. Phys. **136**(9), 094506 (2012).

[18] A.D. Fortes, I.G. Wood, M. Alfredsson, L. Vočadlo, K.S. Knight, W.G. Marshall, M.G. Tucker, and F. Fernandez-Alonso, "The high-pressure phase diagram of ammonia dihydrate," High Press. Res. **27**(2), 201–212 (2007).

[19] X. Li, W. Shi, X. Liu, and Z. Mao, "High-pressure phase stability and elasticity of ammonia hydrate," American Mineralogist **104**(9), 1307–1314 (2019).

[20] A. Mondal, K. Mohrbach, T. Fedotenko, M. Bethkenhagen, H.-P. Liermann, and C. Sanchez-Valle, "Evidence for ultra-water-rich ammonia hydrates stabilized in icy exoplanetary mantles," arXiv:2508.11249.

[21] G.W. Lee, W.J. Evans, and C.-S. Yoo, "Crystallization of water in a dynamic diamond-anvil cell: Evidence for ice VII-like local order in supercompressed water," Phys. Rev. B **74**(13), 134112 (2006).

[22] A. Howard, M. Kim, J. Smith, and C.-S. Yoo, "Dynamic compression effects of H 2 O in a dynamic diamond anvil cell: Origin of metastable ice VII and its crystal growth kinetics," Phys. Rev. B **111**(10), 104105 (2025).

[23] Zs. Jenei, H.P. Liermann, R. Husband, A.S.J. Méndez, D. Pennicard, H. Marquardt, E.F. O'Bannon, A. Pakhomova, Z. Konopkova, K. Glazyrin, M. Wendt, S. Wenz, E.E. McBride, W. Morgenroth, B. Winkler, A. Rothkirch, M. Hanfland, and W.J. Evans, "New dynamic diamond anvil cells for tera-pascal per second fast compression x-ray diffraction experiments," Review of Scientific Instruments **90**(6), 065114 (2019).

[24] H.-P. Liermann, Z. Konôpková, W. Morgenroth, K. Glazyrin, J. Bednarčik, E.E. McBride, S. Petitgirard, J.T. Delitz, M. Wendt, Y. Bican, A. Ehnes, I. Schwark, A. Rothkirch, M. Tischer, J. Heuer, H. Schulte-Schrepping, T. Kracht, and H. Franz, "The Extreme Conditions Beamline P02.2 and the Extreme Conditions Science Infrastructure at PETRA III," J. Synchrotron. Rad. **22**(4), 908–924 (2015).

[25] D. Pennicard, S. Lange, S. Smoljanin, H. Hirsemann, H. Graafsma, M. Epple, M. Zuvic, M.-O. Lampert, T. Fritzsch, and M. Rothermund, "The LAMBDA photon-counting pixel detector," J. Phys.: Conf. Ser. **425**(6), 062010 (2013).

[26] D. Pennicard, S. Smoljanin, F. Pithan, M. Sarajlic, A. Rothkirch, Y. Yu, H.P. Liermann, W. Morgenroth, B. Winkler, Z. Jenei, H. Stawitz, J. Becker, and H. Graafsma, "LAMBDA 2M GaAs—A multi-megapixel hard X-ray detector for synchrotrons," J. Inst. **13**(01), C01026–C01026 (2018).

[27] C. Prescher, and V.B. Prakapenka, "*DIOPTAS* : a program for reduction of two-dimensional X-ray diffraction data and data exploration," High Press. Res. **35**(3), 223–230 (2015).

[28] M. Karnevskiy, K. Glazyrin, Y. Yu, A. Mondal, C. Sanchez-Valle, H. Marquardt, R. Husband, E. O'Bannon, C. Prescher, A. Barty, and H.-P. Liermann, "Automated pipeline processing X-ray diffraction data from dynamic compression experiments on the Extreme Conditions Beamline of PETRA III," J. Appl. Crystallogr. **57**(4), 1217–1228 (2024).

[29] A. Mondal, R.J. Husband, H.-P. Liermann, and C. Sanchez-Valle, "Equation of state of ammonia dihydrate up to 112 GPa by static and dynamic compression experiments in diamond anvil cells," Phys. Rev. B **107**(22), 224108 (2023).

[30] Y. Fei, A. Ricolleau, M. Frank, K. Mibe, G. Shen, and V. Prakapenka, "Toward an internally consistent pressure scale," Proc. Natl. Acad. Sci. U.S.A. **104**(22), 9182–9186 (2007).

[31] K. Takemura, and A. Dewaele, "Isothermal equation of state for gold with a He-pressure medium," Phys. Rev. B **78**(10), 104119 (2008).

[32] J. Gonzalez-Platas, M. Alvaro, F. Nestola, and R. Angel, "*EosFit7-GUI*: a new graphical user interface for equation of state calculations, analyses and teaching," J Appl Crystallogr **49**(4), 1377–1382 (2016).

[33] C. Ma, F. Li, Q. Zhou, F. Huang, J. Wang, M. Zhang, Z. Wang, and Q. Cui, "Ammonia molecule rotation of pressure-induced phase transition in ammonia hemihydrates 2NH3·H2O," RSC Adv. **2**(11), 4920 (2012).

[34] L. Andriambariarijaona, F. Datchi, H. Zhang, K. Béneut, B. Baptiste, N. Guignot, and S. Ninet, "High pressure–temperature phase diagram of ammonia hemihydrate," Phys. Rev. B **108**(17), 174102 (2023).

[35] W. Xu, V.N. Robinson, X. Zhang, H.-C. Zhang, M.-E. Donnelly, P. Dalladay-Simpson, A. Hermann, X.-D. Liu, and E. Gregoryanz, "Ionic Phases of Ammonia-Rich Hydrate at High Densities," Phys. Rev. Lett. **126**(1), 015702 (2021).

[36] C. Liu, A. Mafety, J.A. Queyroux, C.W. Wilson, H. Zhang, K. Béneut, G. Le Marchand, B. Baptiste, P. Dumas, G. Garbarino, F. Finocchi, J.S. Loveday, F. Pietrucci, A.M. Saitta, F. Datchi, and S. Ninet, "Topologically frustrated ionisation in a water-ammonia ice mixture," Nat. Commun. **8**(1), 1065 (2017).

[37] H. Zhang, F. Datchi, L. Andriambariarijaona, M. Rescigno, L.E. Bove, S. Klotz, and S. Ninet, "Observation of a Plastic Crystal in Water–Ammonia Mixtures under High Pressure and Temperature," J. Phys. Chem. Lett. **14**(9), 2301–2307 (2023).

[38] H. Zhang, Experimental investigation of the phase diagram of ammonia monohydrate at high pressure and temperature, Ph.D. thesis, HAL Sorbonne Université, 2019, https://theses.hal.science/tel-03459424.

[39] J.S. Loveday, and R.J. Nelmes, "Ammonia Monohydrate VI: A Hydrogen-Bonded Molecular Alloy," Phys. Rev. Lett. **83**(21), 4329–4332 (1999).

[40] S. Klotz, K. Komatsu, H. Kagi, K. Kunc, A. Sano-Furukawa, S. Machida, and T. Hattori, "Bulk moduli and equations of state of ice VII and ice VIII," Phys. Rev. B **95**(17), 174111 (2017).

[41] F. Datchi, S. Ninet, M. Gauthier, A.M. Saitta, B. Canny, and F. Decremps, "Solid ammonia at high pressure: A single-crystal x-ray diffraction study to 123 GPa," Phys. Rev. B **73**(17), 174111 (2006).

[42] S. Berni, D. Scelta, S. Fanetti, and R. Bini, "High pressure behavior of ethylene and water: From clathrate hydrate to polymerization in solid ice mixtures," The Journal of Chemical Physics **158**(6), 064505 (2023).

[43] X. Wei, X. Li, Z. Zhang, H.-P. Liermann, S. Speziale, M. Pu, C. Zhang, R. Li, H. Yu, L. Li, F. Li, and Q. Zhou, "Impact of CaC l 2 -induced chemical pressure on the phase transition of H 2 O at high pressure," Phys. Rev. B **109**(13), 134108 (2024).

[44] A.K. Singh, and T. Kenichi, "Measurement and analysis of nonhydrostatic lattice strain component in niobium to 145 GPa under various fluid pressure-transmitting media," Journal of Applied Physics **90**(7), 3269–3275 (2001).

[45] H. Zhang, F. Datchi, L.M. Andriambariarijaona, G. Zhang, J.A. Queyroux, K. Béneut, M. Mezouar, and S. Ninet, "Melting curve and phase diagram of ammonia monohydrate at high pressure and temperature," J. Chem. Phys. **153**(15), 154503 (2020).

[46] A.S.J. Méndez, F. Trybel, R.J. Husband, G. Steinle-Neumann, H.-P. Liermann, and H. Marquardt, "Bulk modulus of H 2 O across the ice VII–ice X transition measured by time-resolved x-ray diffraction in dynamic diamond anvil cell experiments," Phys. Rev. B **103**(6), 064104 (2021).

[47] B. Journaux, I. Daniel, S. Petitgirard, H. Cardon, J.-P. Perrillat, R. Caracas, and M. Mezouar, "Salt partitioning between water and high-pressure ices. Implication for the dynamics and habitability of icy moons and water-rich planetary bodies," Earth and Planetary Science Letters **463**, 36–47 (2017).

[48] M.R. Frank, C.E. Runge, H.P. Scott, S.J. Maglio, J. Olson, V.B. Prakapenka, and G. Shen, "Experimental study of the NaCl–H2O system up to 28GPa: Implications for ice-rich planetary bodies," Physics of the Earth and Planetary Interiors **155**(1–2), 152–162 (2006).

[49] M.R. Frank, E. Aarestad, H.P. Scott, and V.B. Prakapenka, "A comparison of ice VII formed in the H2O, NaCl–H2O, and CH3OH–H2O systems: Implications for H2O-rich planets," Physics of the Earth and Planetary Interiors **215**, 12–20 (2013).

# Figures

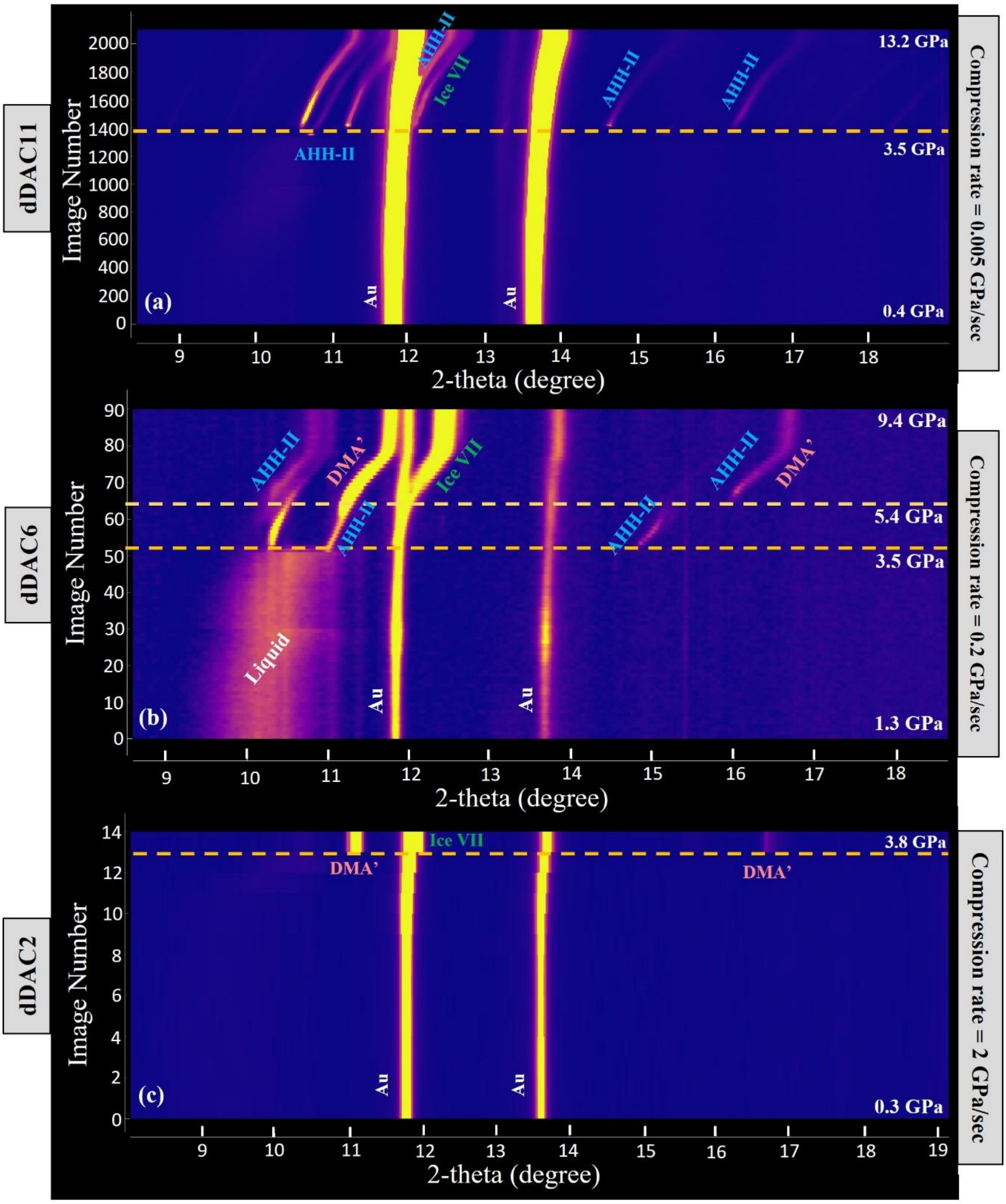

**Figure 1:** Intensity plot displaying the stacking of all diffractograms collected in (a) dDAC11 (0.005 GPa/sec, **Table I**), (b) dDAC6 (0.2 GPa/sec, **Table I**) and (c) dDAC2 (2 GPa/sec, **Table I**) during ramp compressions. The 25 wt% $NH_3$-$H_2O$ solutions crystallize into either (a) AHH-II phase or (b) the bcc DMA' phase or (c) a mixture of both assemblages depending on the compression rate. The pressure values indicate the initial and highest pressure in the ramp compressions and highlight the crystallization of different phases observed with increasing

pressure from the liquid phase. In the dDAC6 run with intermediate compression rate, we were only able to identify the DMA’ phase from ~5.4 GPa onwards, due to the overlap of the DMA’ (110) reflection with the (023) AHH-II reflection.

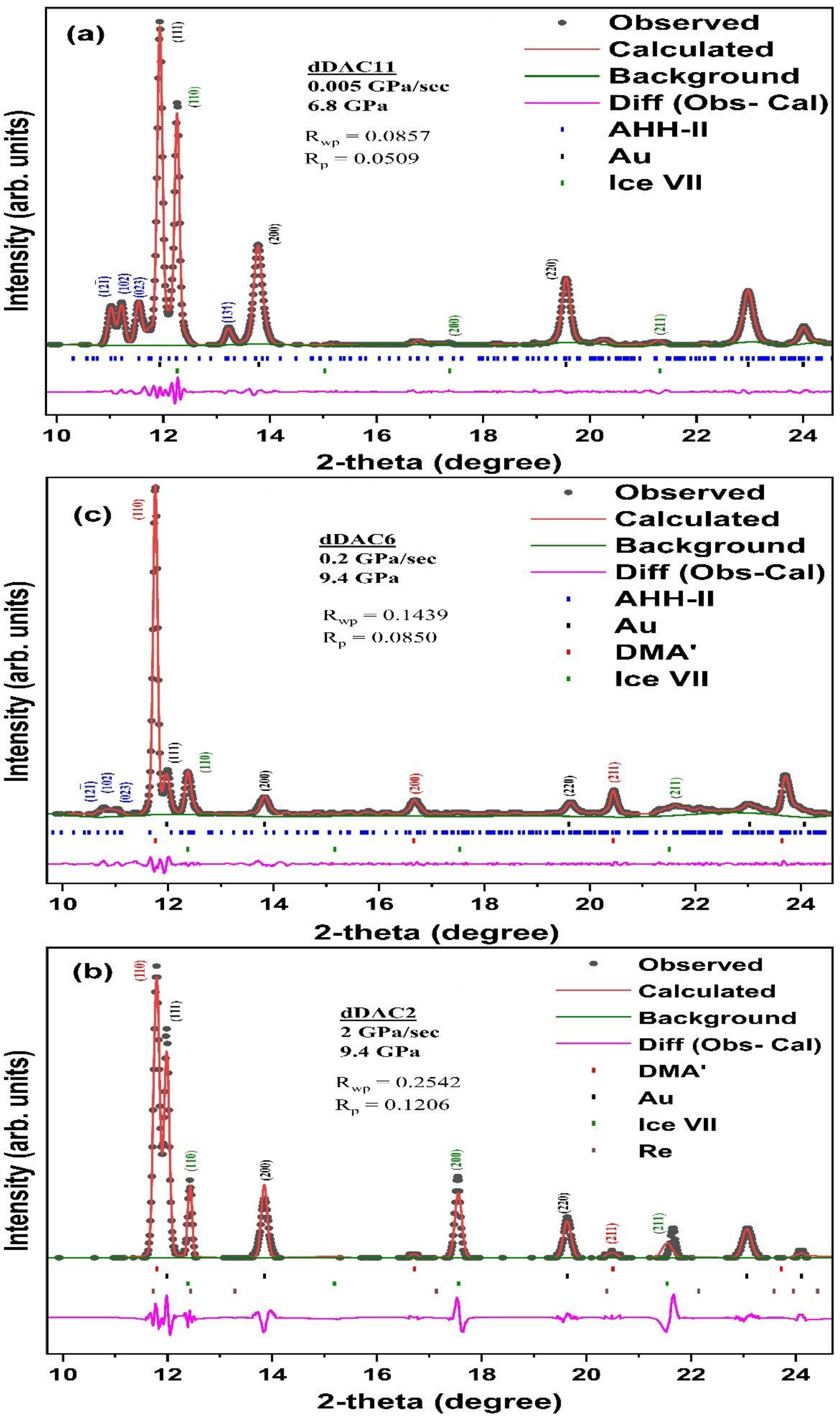

**Figure 2.** Le Bail refinements of selected 1-dimensional (1D) x-ray diffraction patterns (X-ray energy = 25.63 keV) displaying the different phase assemblages crystallized from the 25 wt% NH3-H2O solution at the indicated compression rates (**Table I**).

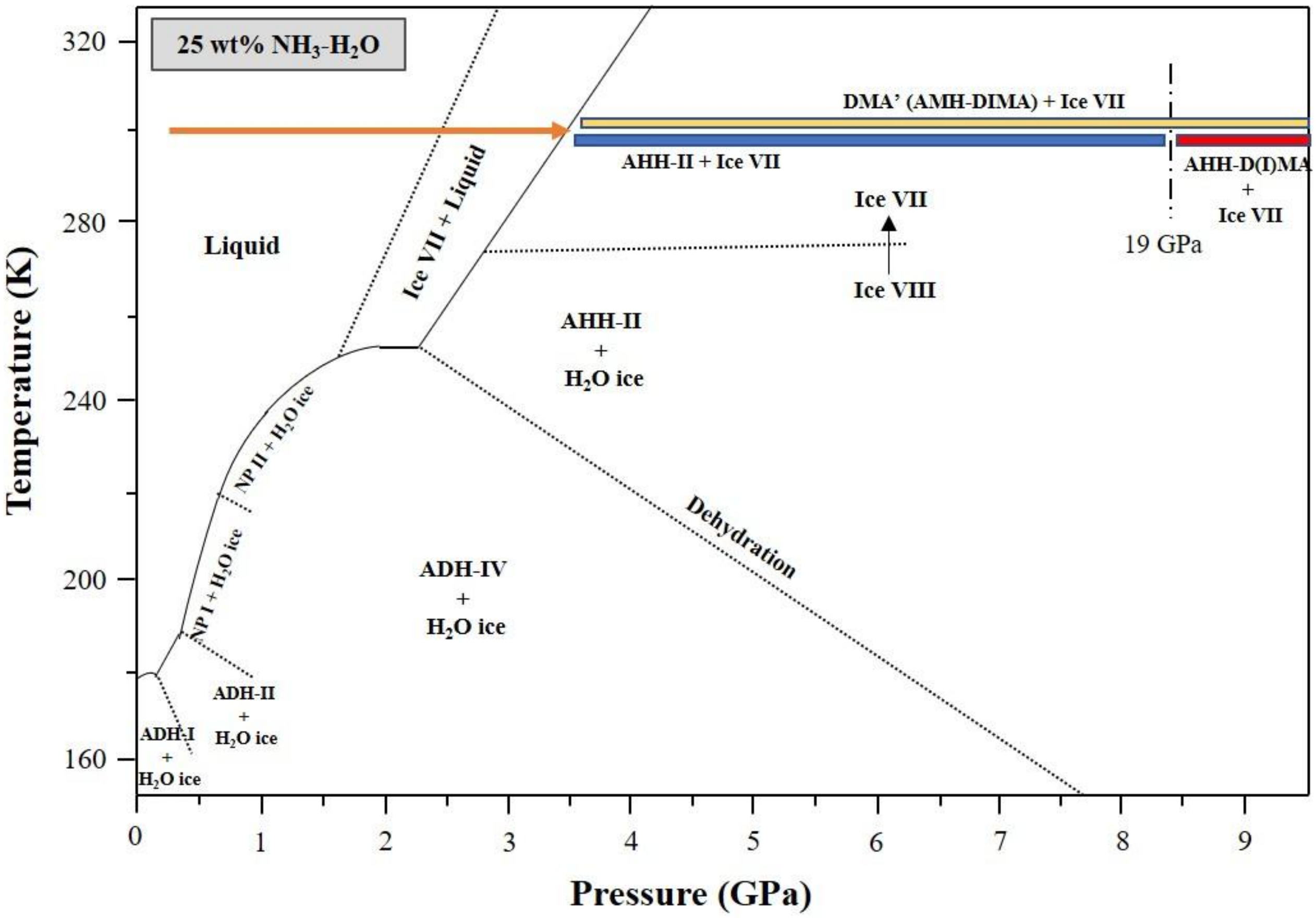

**Figure 3.** Pressure-temperature phase diagram of 25 wt% $NH_3$-$H_2O$ system modified after Mousis et al.[13] and Wilson et al.[14] based on the results of the present compression-rate dependent crystallization studies. The solution crystallizes to either AHH-II phase + $H_2O$ ice VII (color coded in blue) or the DMA' phase (color coded in yellow) or in a mixture of the two assemblages. The new phase (NP)-I and II are taken from previous study by Mousis et al.[13] on 25 wt% $NH_3$-$H_2O$ system, the structures of which are yet to be constrained.

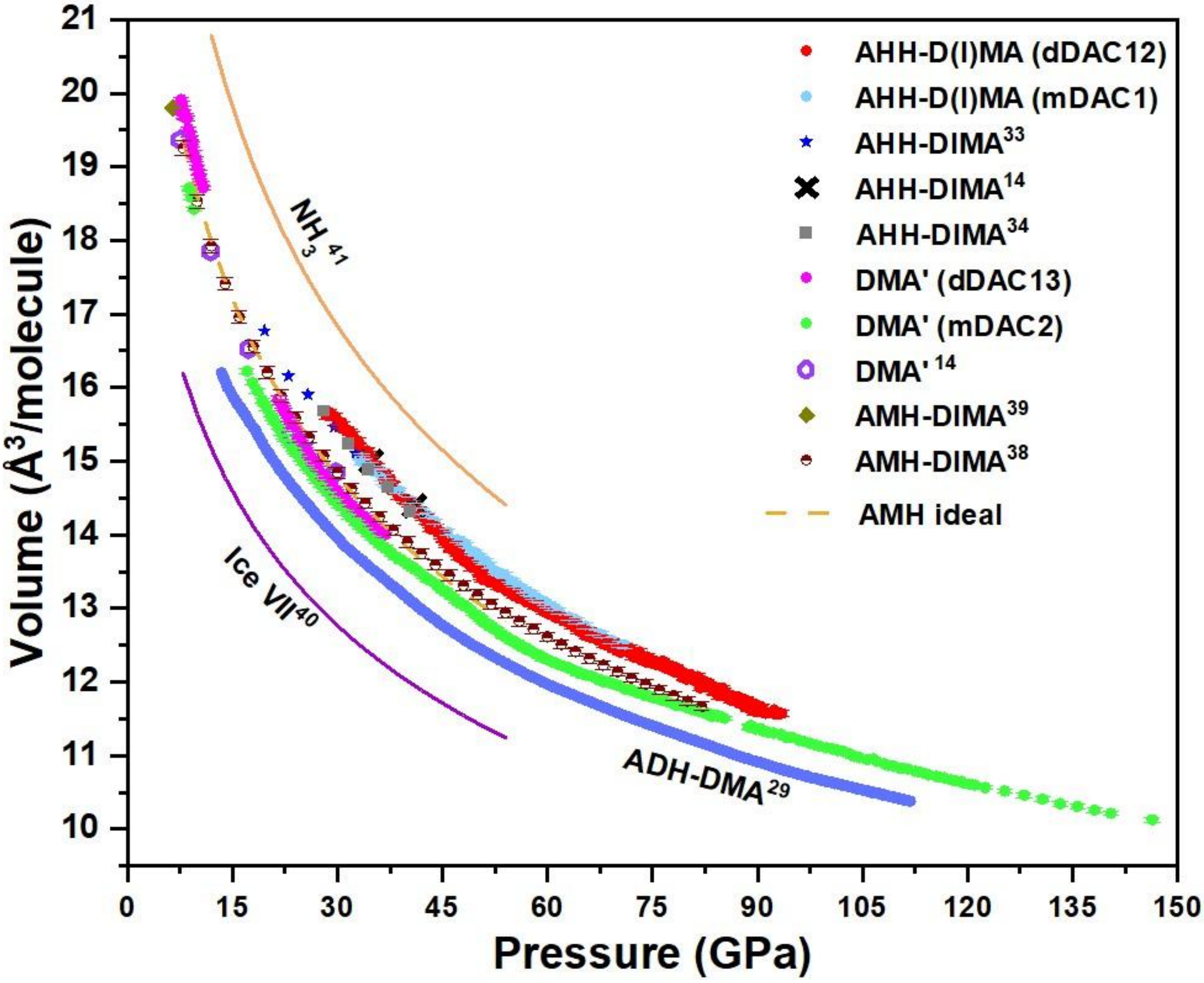

**Figure 4.** Comparison of pressure-volume data between AHH-D(I)MA and DMA' phases (circles), derived from the evolution of the (110) reflection in all four compression experiments listed in **Table II**. The available compression data for DMA/DIMA phases within the $NH_3$-$H_2O$ system, including pure $NH_3$[41] and $H_2O$ ice VII[40] are reported for comparison. The pressure-volume data of DMA' phase align closely with data reported for AMH-DIMA[38] and calculated for an hypothetical 'AMH ideal' phase (i.e. ideal mixing in volume between $NH_3$ and $H_2O$ with 1:1 ratio) considering their mutual uncertainties.

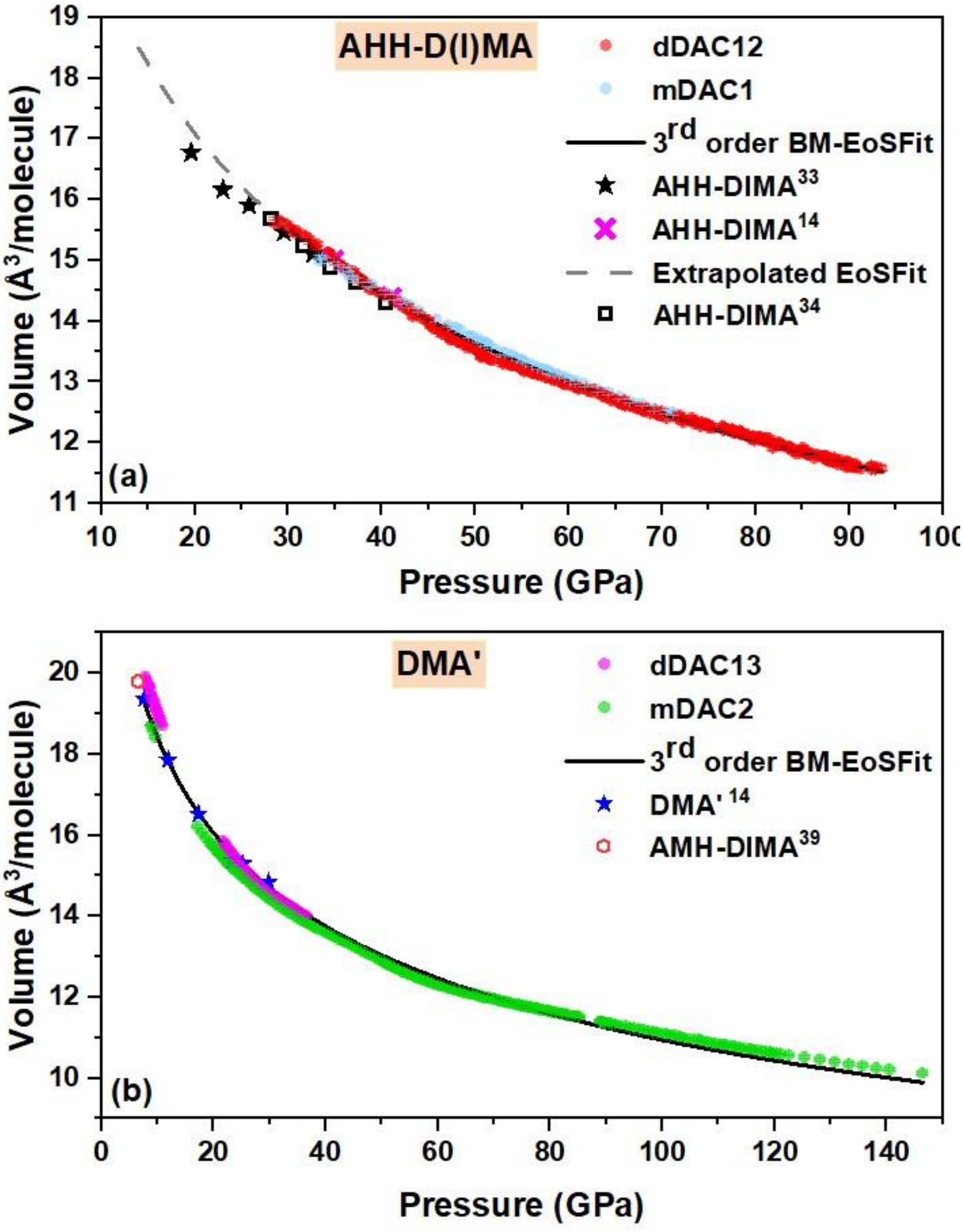

**Figure 5. (a)** Pressure-volume data from static (mDAC) and dynamic (dDAC) compression runs (**Table II**) on the AHH-D(I)MA phase up to 93 GPa (circles) and a 3$^{rd}$ order Birch-Murnaghan equation of state fit to the data (solid black line). The P-V data along with the extrapolated EoSFit (dashed grey line), aligns well with the reported literature values. **(b)** Pressure-volume data of the DMA' phase (circles) to 146 GPa from both sDAC and dDAC compression experiments (**Table II**) and the 3$^{rd}$ order Birch-Murnaghan EoS fit to the data (solid black line) together with literature data.

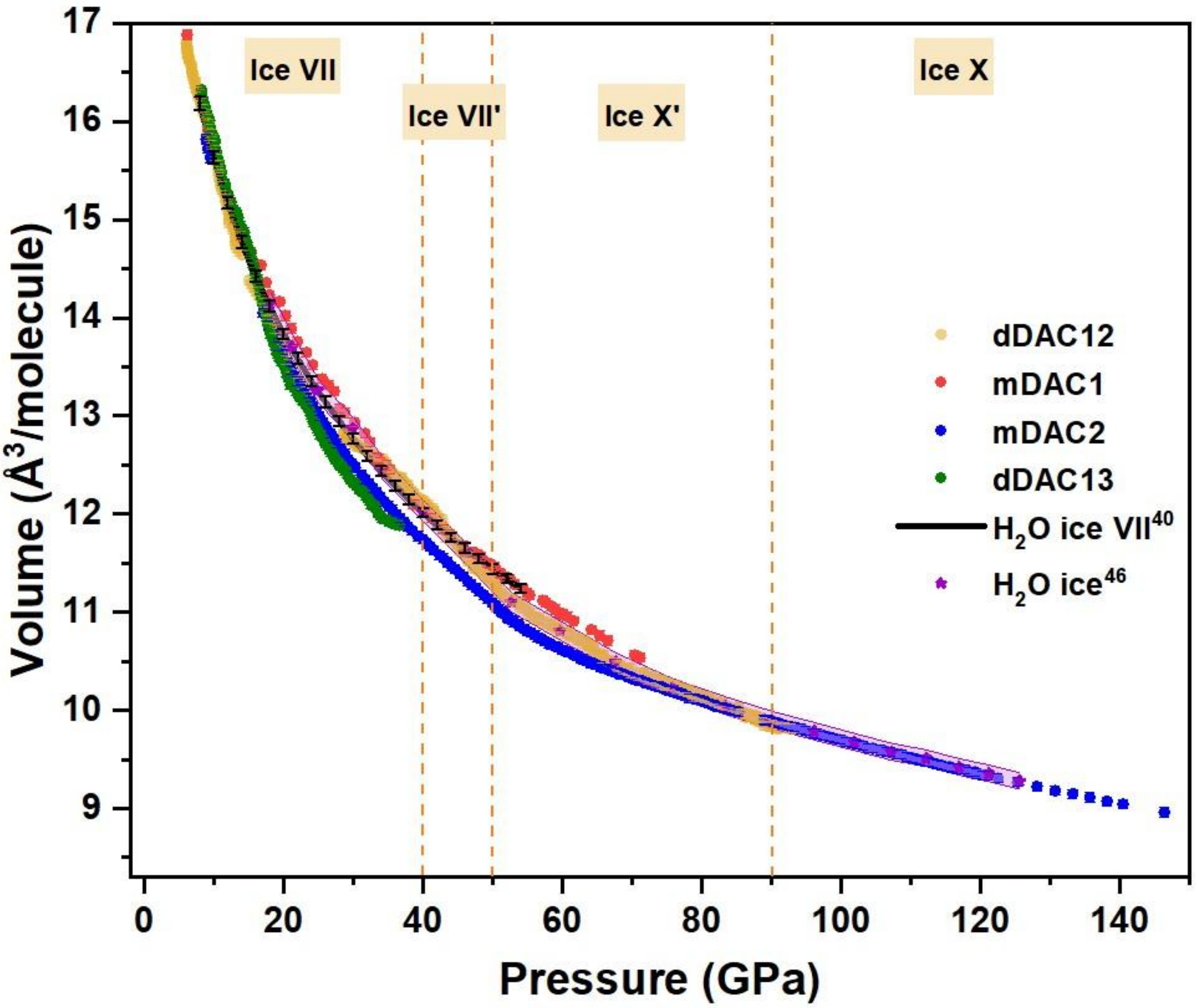

**Figure 6.** Pressure-volume data of $H_2O$ ice coexisting with the ammonia hydrates (AHH-II/AHH-D(I)MA and DMA') in four different compression runs (**Table II**) and comparison with literature data[40,46]. The vertical dashed lines indicate the stability field of the different $H_2O$ ice polymorphs[46]. The negative volume deviations above 30 GPa may be related to the incorporation of $NH_3$ impurities in the ice VII lattice. Above 60 GPa, $H_2O$ ice volume data from all four compression runs tend to match again suggesting a more ideal behavior at higher pressures.

# Compressional rate-dependent stability of ammonia hydrates upon crystallization of water-rich ammonia-water solutions

A. Mondal[1,2 *], K. Mohrbach[1,3], K. Glazyrin[3], H.-P. Liermann[3], C. Sanchez-Valle[1]

[1] Institut für Mineralogie, Universität Münster, 48149 Münster, Germany

[2] Department of Physics, NIT Agarata, 799046 Jirania, India

[3] Deutsches Elektronen-Synchrotron DESY, Notkestr. 85, 22607 Hamburg, Germany

*** Author to whom correspondence should be addressed:** anshuman.phy@faculty.nita.ac.in

# Supplementary information

## Content of this file

This file contains additional figures (Figure S1, S2 and S3) and tables (Table S1 and S2).

**Figure S1:** Pressure evolution of XRD patterns collected in mDAC2 and dDAC13 runs.

**Figure S2:** Pressure evolution of XRD patterns collected in mDAC2 and dDAC13 runs.

**Figure S3:** X-ray diffraction maps at selected dDAC compression runs.

**Table S1:** Calculated P-V compression data of AHH-D(I)MA and DMA' phases.

**Table S2:** Calculated P-V compression data of $H_2O$ ice phases.

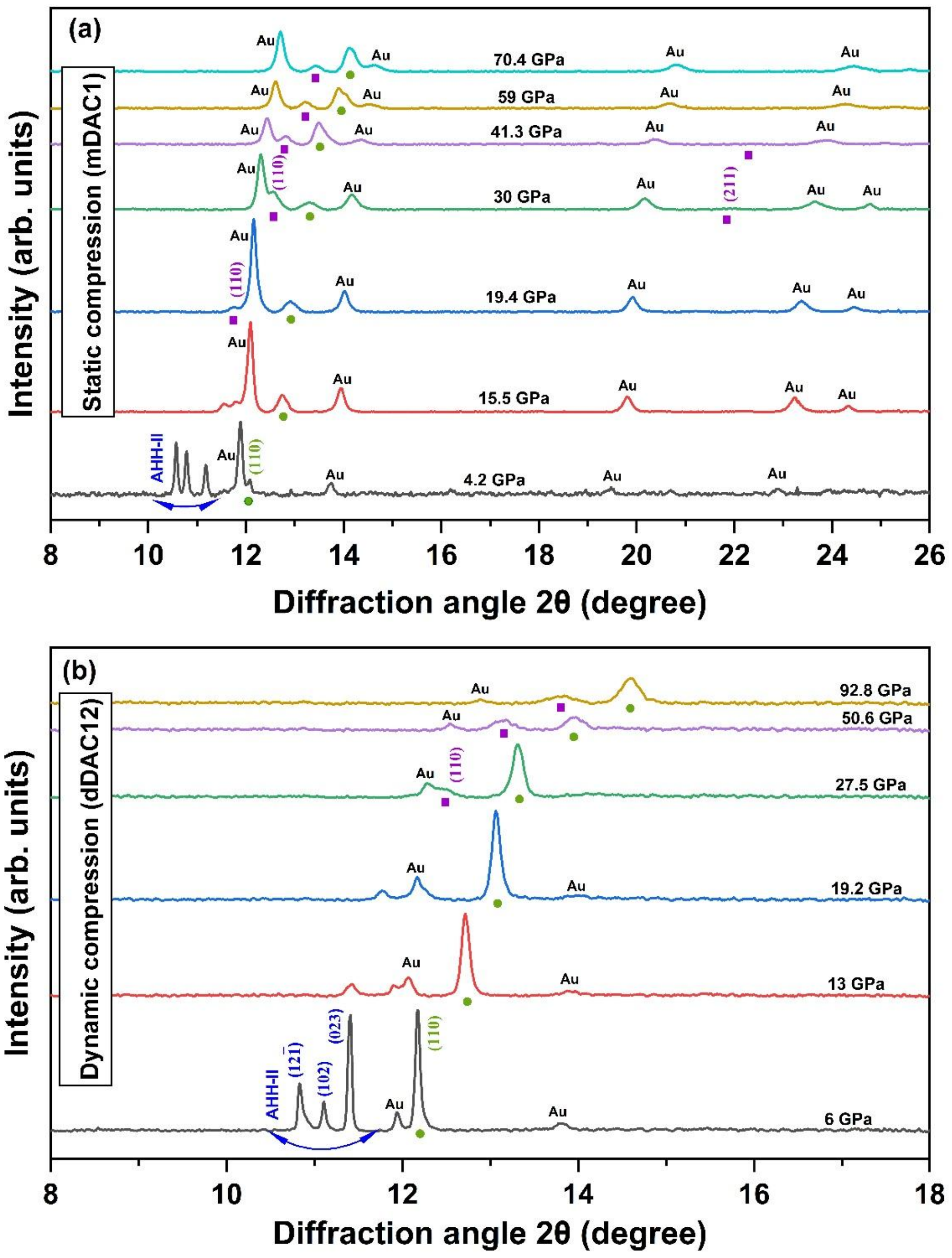

**Figure S1.** Selected XRD patterns documenting the pressure evolution of AHH-II and $H_2O$ ice phases, recorded at 298 K from **(a)** membrane-driven compression mDAC1 and **(b)** dynamic compression dDAC12 experiments (**Table II**), respectively. The distinctive Bragg reflections of AHH-II phase are indicated by blue curved arrows, while those of $H_2O$ ice VII and AHH-D(I)MA are marked by green circles and purple squares, respectively. The remaining major reflections correspond to the gold (Au) pressure marker.

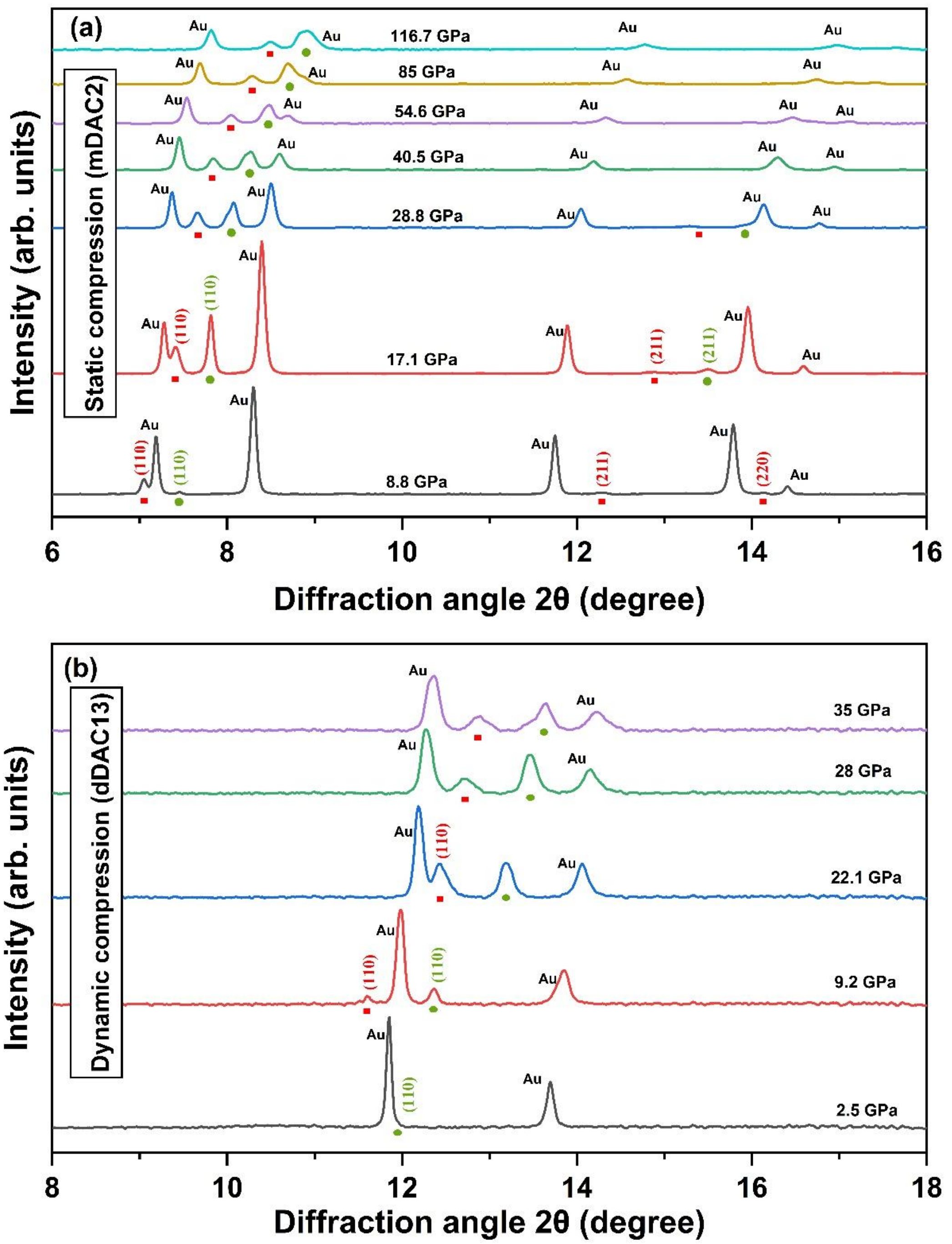

**Figure S2.** Selected XRD patterns illustrating the pressure evolution of the DMA' and $H_2O$ ice phases, recorded at 298 K from **(a)** static compression mDAC2 (incident x-ray energy = 42.65 keV) and **(b)** dynamic compression dDAC13 (incident x-ray energy = 25.62 keV) experiments (**Table II**). The Bragg reflections corresponding to $H_2O$ ice and DMA' phases are marked by green circles and red squares respectively. The bcc DMA' phase is stable up to the highest pressure investigated here. The other prominent reflections correspond to the Au pressure marker.

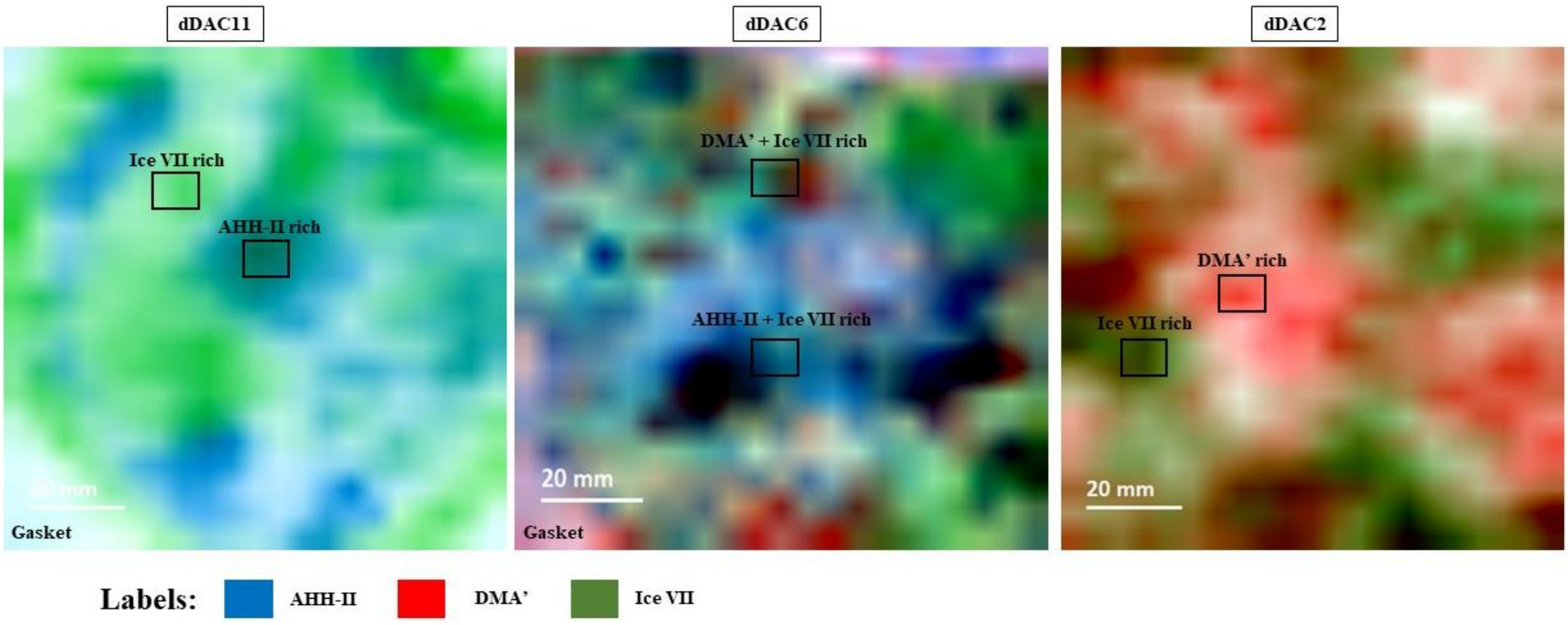

**Figure S3.** X-ray diffraction maps displaying distribution of AHH-II, DMA' and Ice VII phases within the sample chamber at the highest-pressure plateau in different dDAC compression runs (**Table I**). The maps covering area of 105 x 105 $\mu m^2$ with 5 µm step-size.

**Table S1:** Calculated pressure-volume (P-V) data of AHH-D(I)MA and DMA’ phases recorded in four different compression runs presented in Figure 4.

| AHH-D(I)MA | | | | | | DMA’ | | | | | |
|---|---|---|---|---|---|---|---|---|---|---|---|
| dDAC12 | | | mDAC1 | | | dDAC13 | | | mDAC2 | | |
| P (GPa) | V (Å$^3$) | Err (Å) | P (GPa) | V (Å$^3$) | Err (Å) | P (GPa) | V (Å$^3$) | Err (Å) | P (GPa) | V (Å$^3$) | Err (Å) |
| 28.52 | 15.62 | 0.04 | 33.37 | 15.00 | 0.04 | 7.71 | 19.91 | 0.03 | 8.85 | 18.71 | 0.03 |
| 28.49 | 15.67 | 0.04 | 33.33 | 15.05 | 0.04 | 7.82 | 19.86 | 0.03 | 8.93 | 18.71 | 0.03 |
| 28.92 | 15.66 | 0.04 | 36.49 | 14.85 | 0.04 | 7.93 | 19.79 | 0.03 | 9.18 | 18.58 | 0.03 |
| 29.29 | 15.62 | 0.04 | 33.51 | 15.05 | 0.04 | 8.04 | 19.74 | 0.03 | 9.54 | 18.42 | 0.03 |
| 28.93 | 15.63 | 0.04 | 33.62 | 15.03 | 0.04 | 8.14 | 19.69 | 0.03 | 17.10 | 16.22 | 0.03 |
| 28.99 | 15.62 | 0.04 | 33.87 | 15.00 | 0.04 | 8.26 | 19.67 | 0.03 | 17.89 | 16.07 | 0.03 |
| 29.67 | 15.60 | 0.04 | 34.82 | 14.94 | 0.04 | 8.39 | 19.69 | 0.03 | 18.54 | 15.95 | 0.03 |
| 30.05 | 15.54 | 0.04 | 35.33 | 14.89 | 0.04 | 8.55 | 19.66 | 0.03 | 19.27 | 15.82 | 0.03 |
| 30.19 | 15.53 | 0.04 | 35.91 | 14.84 | 0.04 | 8.71 | 19.52 | 0.03 | 19.60 | 15.78 | 0.03 |
| 30.74 | 15.50 | 0.04 | 36.48 | 14.77 | 0.04 | 8.85 | 19.44 | 0.03 | 19.89 | 15.72 | 0.03 |
| 30.42 | 15.49 | 0.04 | 36.91 | 14.72 | 0.04 | 9.01 | 19.38 | 0.03 | 20.48 | 15.63 | 0.03 |
| 30.79 | 15.45 | 0.04 | 36.74 | 14.72 | 0.04 | 9.17 | 19.34 | 0.03 | 20.77 | 15.59 | 0.03 |
| 31.36 | 15.43 | 0.04 | 36.91 | 14.69 | 0.04 | 9.32 | 19.31 | 0.03 | 21.12 | 15.53 | 0.03 |
| 31.82 | 15.40 | 0.04 | 37.28 | 14.67 | 0.04 | 9.50 | 19.18 | 0.03 | 21.57 | 15.45 | 0.03 |
| 32.19 | 15.36 | 0.04 | 38.62 | 14.61 | 0.04 | 9.68 | 19.11 | 0.03 | 22.04 | 15.39 | 0.03 |
| 32.34 | 15.30 | 0.04 | 38.91 | 14.60 | 0.04 | 9.85 | 19.10 | 0.03 | 22.40 | 15.33 | 0.03 |
| 33.00 | 15.27 | 0.04 | 40.36 | 14.47 | 0.04 | 10.04 | 19.00 | 0.03 | 22.55 | 15.30 | 0.03 |
| 32.71 | 15.25 | 0.04 | 41.30 | 14.40 | 0.04 | 10.22 | 18.92 | 0.03 | 22.99 | 15.25 | 0.03 |
| 33.02 | 15.21 | 0.04 | 42.11 | 14.32 | 0.04 | 10.39 | 18.87 | 0.03 | 23.36 | 15.21 | 0.03 |
| 34.21 | 15.13 | 0.04 | 42.89 | 14.26 | 0.04 | 10.59 | 18.78 | 0.03 | 23.69 | 15.16 | 0.03 |
| 34.26 | 15.14 | 0.04 | 43.92 | 14.18 | 0.04 | 10.80 | 18.71 | 0.03 | 23.88 | 15.14 | 0.03 |
| 34.47 | 15.10 | 0.04 | 45.78 | 14.03 | 0.04 | 21.63 | 15.85 | 0.03 | 24.28 | 15.09 | 0.03 |
| 34.14 | 15.07 | 0.04 | 47.08 | 13.93 | 0.04 | 21.86 | 15.80 | 0.03 | 24.70 | 15.03 | 0.03 |
| 34.63 | 15.06 | 0.04 | 47.91 | 13.89 | 0.04 | 22.10 | 15.75 | 0.03 | 25.16 | 14.97 | 0.03 |
| 34.78 | 15.02 | 0.04 | 48.35 | 13.86 | 0.04 | 22.31 | 15.72 | 0.03 | 25.65 | 14.92 | 0.03 |
| 34.85 | 15.00 | 0.04 | 49.25 | 13.76 | 0.04 | 22.49 | 15.68 | 0.03 | 26.18 | 14.86 | 0.03 |
| 35.12 | 14.96 | 0.04 | 49.60 | 13.76 | 0.04 | 22.72 | 15.64 | 0.03 | 26.78 | 14.78 | 0.03 |
| 35.44 | 14.94 | 0.04 | 50.28 | 13.71 | 0.04 | 22.93 | 15.61 | 0.03 | 27.21 | 14.73 | 0.03 |
| 35.65 | 14.92 | 0.04 | 50.50 | 13.66 | 0.04 | 23.13 | 15.56 | 0.03 | 27.66 | 14.67 | 0.03 |
| 35.79 | 14.88 | 0.04 | 50.97 | 13.63 | 0.04 | 23.36 | 15.52 | 0.03 | 28.25 | 14.61 | 0.03 |
| 36.22 | 14.86 | 0.04 | 51.28 | 13.63 | 0.04 | 23.58 | 15.48 | 0.03 | 28.80 | 14.56 | 0.03 |
| 36.93 | 14.82 | 0.04 | 51.73 | 13.58 | 0.04 | 23.78 | 15.44 | 0.03 | 29.31 | 14.49 | 0.03 |
| 36.99 | 14.78 | 0.04 | 52.53 | 13.50 | 0.04 | 23.94 | 15.40 | 0.03 | 29.85 | 14.44 | 0.03 |
| 36.89 | 14.76 | 0.04 | 53.22 | 13.48 | 0.04 | 24.15 | 15.36 | 0.03 | 30.28 | 14.39 | 0.03 |
| 37.10 | 14.75 | 0.04 | 53.93 | 13.45 | 0.04 | 24.35 | 15.33 | 0.03 | 31.22 | 14.30 | 0.03 |
| 37.19 | 14.74 | 0.04 | 54.46 | 13.39 | 0.04 | 24.58 | 15.29 | 0.03 | 31.69 | 14.25 | 0.03 |
| 37.78 | 14.70 | 0.04 | 55.01 | 13.36 | 0.04 | 24.82 | 15.26 | 0.03 | 32.34 | 14.20 | 0.03 |
| 37.93 | 14.67 | 0.04 | 55.31 | 13.33 | 0.04 | 25.04 | 15.22 | 0.03 | 32.80 | 14.16 | 0.03 |
| 37.45 | 14.66 | 0.04 | 56.04 | 13.29 | 0.04 | 25.28 | 15.18 | 0.03 | 33.31 | 14.11 | 0.03 |
| 37.92 | 14.67 | 0.04 | 56.63 | 13.24 | 0.04 | 25.56 | 15.15 | 0.03 | 33.87 | 14.06 | 0.03 |
| 38.23 | 14.61 | 0.04 | 57.23 | 13.22 | 0.04 | 25.82 | 15.11 | 0.03 | 34.50 | 14.01 | 0.03 |
| 38.37 | 14.59 | 0.04 | 57.69 | 13.19 | 0.04 | 26.09 | 15.07 | 0.03 | 35.08 | 13.95 | 0.03 |
| 38.64 | 14.58 | 0.04 | 57.91 | 13.17 | 0.04 | 26.32 | 15.04 | 0.03 | 35.63 | 13.91 | 0.03 |
| 38.72 | 14.52 | 0.04 | 58.47 | 13.13 | 0.04 | 26.59 | 15.00 | 0.03 | 36.22 | 13.86 | 0.03 |
| 39.15 | 14.51 | 0.04 | 58.95 | 13.11 | 0.04 | 26.84 | 14.96 | 0.03 | 36.79 | 13.82 | 0.03 |
| 39.69 | 14.47 | 0.04 | 59.58 | 13.08 | 0.04 | 27.08 | 14.93 | 0.03 | 37.33 | 13.78 | 0.03 |
| 40.02 | 14.45 | 0.04 | 60.10 | 13.04 | 0.04 | 27.32 | 14.89 | 0.03 | 37.89 | 13.73 | 0.03 |
| 40.30 | 14.42 | 0.04 | 60.71 | 13.01 | 0.04 | 27.54 | 14.85 | 0.03 | 38.48 | 13.69 | 0.03 |
| 40.81 | 14.42 | 0.04 | 61.67 | 12.96 | 0.04 | 27.77 | 14.82 | 0.03 | 39.32 | 13.64 | 0.03 |
| 40.68 | 14.38 | 0.04 | 63.08 | 12.88 | 0.04 | 28.02 | 14.79 | 0.03 | 40.05 | 13.60 | 0.03 |
| 41.41 | 14.32 | 0.04 | 64.19 | 12.83 | 0.04 | 28.25 | 14.76 | 0.03 | 40.49 | 13.57 | 0.03 |
| 41.54 | 14.33 | 0.04 | 65.38 | 12.76 | 0.04 | 28.56 | 14.73 | 0.03 | 41.05 | 13.53 | 0.03 |
| 41.59 | 14.34 | 0.04 | 66.52 | 12.68 | 0.04 | 28.83 | 14.70 | 0.03 | 41.55 | 13.50 | 0.03 |
| 42.06 | 14.27 | 0.04 | 67.75 | 12.64 | 0.04 | 29.11 | 14.67 | 0.03 | 42.40 | 13.43 | 0.03 |
| 42.27 | 14.24 | 0.04 | 69.08 | 12.56 | 0.04 | 29.37 | 14.64 | 0.03 | 42.95 | 13.39 | 0.03 |
| 42.78 | 14.24 | 0.04 | 70.40 | 12.51 | 0.04 | 29.65 | 14.61 | 0.03 | 43.49 | 13.36 | 0.03 |
| 42.59 | 14.24 | 0.04 | 71.14 | 12.50 | 0.04 | 29.97 | 14.58 | 0.03 | 44.12 | 13.33 | 0.03 |
| 42.82 | 14.19 | 0.04 | 33.37 | 15.00 | 0.04 | 30.24 | 14.55 | 0.03 | 44.65 | 13.28 | 0.03 |
| 43.06 | 14.17 | 0.04 | | | | 30.49 | 14.52 | 0.03 | 45.31 | 13.24 | 0.03 |
| 43.28 | 14.13 | 0.04 | | | | 30.84 | 14.50 | 0.03 | 45.84 | 13.20 | 0.03 |
| 43.60 | 14.12 | 0.04 | | | | 31.17 | 14.47 | 0.03 | 46.41 | 13.16 | 0.03 |
| 44.12 | 14.10 | 0.04 | | | | 31.45 | 14.44 | 0.03 | 46.98 | 13.11 | 0.03 |
| 43.29 | 14.08 | 0.04 | | | | 31.76 | 14.41 | 0.03 | 47.59 | 13.07 | 0.03 |
| 44.19 | 14.05 | 0.04 | | | | 32.05 | 14.38 | 0.03 | 48.09 | 13.04 | 0.03 |
| 44.90 | 14.02 | 0.04 | | | | 32.34 | 14.36 | 0.03 | 48.79 | 12.99 | 0.03 |
| 44.83 | 14.01 | 0.04 | | | | 32.60 | 14.33 | 0.03 | 49.37 | 12.94 | 0.03 |
| 44.99 | 13.98 | 0.04 | | | | 32.94 | 14.31 | 0.03 | 49.93 | 12.90 | 0.03 |
| 45.22 | 13.95 | 0.04 | | | | 33.22 | 14.28 | 0.03 | 50.37 | 12.88 | 0.03 |
| 45.20 | 13.93 | 0.04 | | | | 33.57 | 14.26 | 0.03 | 50.55 | 12.85 | 0.03 |
| 45.56 | 13.90 | 0.04 | | | | 33.89 | 14.23 | 0.03 | 50.86 | 12.82 | 0.03 |
| 45.66 | 13.87 | 0.04 | | | | 34.16 | 14.20 | 0.03 | 51.21 | 12.80 | 0.03 |
| 46.75 | 13.87 | 0.04 | | | | 34.46 | 14.17 | 0.03 | 51.43 | 12.79 | 0.03 |
| 46.25 | 13.83 | 0.04 | | | | 34.79 | 14.14 | 0.03 | 51.69 | 12.77 | 0.03 |
| 46.29 | 13.82 | 0.04 | | | | 35.13 | 14.11 | 0.03 | 52.06 | 12.75 | 0.03 |
| 46.37 | 13.80 | 0.04 | | | | 35.43 | 14.08 | 0.03 | 52.12 | 12.74 | 0.03 |
| 47.27 | 13.78 | 0.04 | | | | 35.75 | 14.06 | 0.03 | 52.38 | 12.73 | 0.03 |
| 47.16 | 13.75 | 0.04 | | | | 36.07 | 14.03 | 0.03 | 52.78 | 12.70 | 0.03 |
| 47.17 | 13.74 | 0.04 | | | | 36.39 | 14.01 | 0.03 | 53.07 | 12.68 | 0.03 |
| 47.44 | 13.71 | 0.04 | | | | 36.65 | 13.99 | 0.03 | 53.65 | 12.63 | 0.03 |
| 47.38 | 13.69 | 0.04 | | | | 7.71 | 19.91 | 0.03 | 53.93 | 12.61 | 0.03 |
| 48.27 | 13.66 | 0.04 | | | | | | | 54.59 | 12.58 | 0.03 |
| 48.13 | 13.66 | 0.04 | | | | | | | 54.98 | 12.55 | 0.03 |
| 48.64 | 13.63 | 0.04 | | | | | | | 55.52 | 12.53 | 0.03 |
| 48.99 | 13.61 | 0.04 | | | | | | | 56.08 | 12.49 | 0.03 |
| 49.35 | 13.60 | 0.04 | | | | | | | 56.48 | 12.48 | 0.03 |
| 49.45 | 13.58 | 0.04 | | | | | | | 56.92 | 12.44 | 0.03 |
| 49.19 | 13.55 | 0.04 | | | | | | | 57.50 | 12.43 | 0.03 |
| 49.88 | 13.55 | 0.04 | | | | | | | 58.10 | 12.40 | 0.03 |
| 50.02 | 13.53 | 0.04 | | | | | | | 58.42 | 12.38 | 0.03 |
| 50.34 | 13.50 | 0.04 | | | | | | | 59.00 | 12.36 | 0.03 |
| 50.43 | 13.49 | 0.04 | | | | | | | 59.41 | 12.32 | 0.03 |
| 50.54 | 13.47 | 0.04 | | | | | | | 60.00 | 12.30 | 0.03 |
| 50.56 | 13.46 | 0.04 | | | | | | | 60.51 | 12.28 | 0.03 |
| 50.62 | 13.43 | 0.04 | | | | | | | 60.93 | 12.27 | 0.03 |
| 51.04 | 13.42 | 0.04 | | | | | | | 61.68 | 12.23 | 0.03 |
| 50.90 | 13.42 | 0.04 | | | | | | | 62.17 | 12.21 | 0.03 |
| 50.63 | 13.40 | 0.04 | | | | | | | 62.64 | 12.20 | 0.03 |
| 51.48 | 13.38 | 0.04 | | | | | | | 62.99 | 12.19 | 0.03 |
| 51.93 | 13.36 | 0.04 | | | | | | | 63.61 | 12.17 | 0.03 |
| 51.62 | 13.36 | 0.04 | | | | | | | 64.01 | 12.15 | 0.03 |
| 52.90 | 13.34 | 0.04 | | | | | | | 64.46 | 12.12 | 0.03 |
| 52.98 | 13.33 | 0.04 | | | | | | | 64.88 | 12.11 | 0.03 |
| 52.89 | 13.30 | 0.04 | | | | | | | 65.43 | 12.09 | 0.03 |
| 53.51 | 13.29 | 0.04 | | | | | | | 66.03 | 12.08 | 0.03 |
| 54.00 | 13.29 | 0.04 | | | | | | | 66.34 | 12.08 | 0.03 |
| 53.96 | 13.27 | 0.04 | | | | | | | 66.78 | 12.06 | 0.03 |
| 53.85 | 13.27 | 0.04 | | | | | | | 67.14 | 12.04 | 0.03 |
| 54.68 | 13.24 | 0.04 | | | | | | | 67.65 | 12.03 | 0.03 |
| 54.98 | 13.23 | 0.04 | | | | | | | 68.00 | 12.00 | 0.03 |
| 54.93 | 13.21 | 0.04 | | | | | | | 68.44 | 12.00 | 0.03 |
| 55.45 | 13.20 | 0.04 | | | | | | | 68.90 | 11.99 | 0.03 |
| 55.24 | 13.17 | 0.04 | | | | | | | 69.31 | 11.98 | 0.03 |
| 55.93 | 13.16 | 0.04 | | | | | | | 69.69 | 11.97 | 0.03 |
| 56.04 | 13.14 | 0.04 | | | | | | | 70.08 | 11.95 | 0.03 |
| 56.59 | 13.13 | 0.04 | | | | | | | 70.54 | 11.93 | 0.03 |
| 56.61 | 13.13 | 0.04 | | | | | | | 70.98 | 11.91 | 0.03 |
| 56.69 | 13.12 | 0.04 | | | | | | | 71.38 | 11.90 | 0.03 |
| 57.09 | 13.11 | 0.04 | | | | | | | 71.91 | 11.89 | 0.03 |
| 57.50 | 13.07 | 0.04 | | | | | | | 72.17 | 11.88 | 0.03 |
| 57.76 | 13.06 | 0.04 | | | | | | | 72.84 | 11.86 | 0.03 |
| 58.40 | 13.06 | 0.04 | | | | | | | 73.32 | 11.84 | 0.03 |
| 58.40 | 13.04 | 0.04 | | | | | | | 73.82 | 11.82 | 0.03 |
| 58.36 | 13.03 | 0.04 | | | | | | | 74.22 | 11.82 | 0.03 |
| 58.91 | 13.02 | 0.04 | | | | | | | 74.63 | 11.80 | 0.03 |
| 59.34 | 13.01 | 0.04 | | | | | | | 74.97 | 11.79 | 0.03 |
| 59.25 | 12.99 | 0.04 | | | | | | | 75.30 | 11.78 | 0.03 |
| 59.78 | 12.96 | 0.04 | | | | | | | 75.81 | 11.77 | 0.03 |
| 59.48 | 12.96 | 0.04 | | | | | | | 76.20 | 11.76 | 0.03 |
| 60.72 | 12.94 | 0.04 | | | | | | | 76.50 | 11.75 | 0.03 |
| 60.52 | 12.92 | 0.04 | | | | | | | 77.06 | 11.73 | 0.03 |
| 60.81 | 12.91 | 0.04 | | | | | | | 77.23 | 11.73 | 0.03 |
| 61.17 | 12.89 | 0.04 | | | | | | | 77.66 | 11.72 | 0.03 |
| 61.04 | 12.88 | 0.04 | | | | | | | 77.81 | 11.71 | 0.03 |
| 61.91 | 12.89 | 0.04 | | | | | | | 78.18 | 11.69 | 0.03 |
| 62.68 | 12.87 | 0.04 | | | | | | | 78.62 | 11.68 | 0.03 |
| 62.28 | 12.84 | 0.04 | | | | | | | 79.06 | 11.68 | 0.03 |
| 62.12 | 12.85 | 0.04 | | | | | | | 79.44 | 11.67 | 0.03 |
| 63.27 | 12.83 | 0.04 | | | | | | | 79.70 | 11.66 | 0.03 |
| 63.59 | 12.82 | 0.04 | | | | | | | 80.19 | 11.65 | 0.03 |
| 62.82 | 12.81 | 0.04 | | | | | | | 80.53 | 11.63 | 0.03 |
| 63.49 | 12.81 | 0.04 | | | | | | | 81.00 | 11.62 | 0.03 |
| 63.80 | 12.78 | 0.04 | | | | | | | 81.63 | 11.60 | 0.03 |
| 63.65 | 12.76 | 0.04 | | | | | | | 81.96 | 11.59 | 0.03 |
| 64.46 | 12.74 | 0.04 | | | | | | | 82.31 | 11.58 | 0.03 |
| 64.50 | 12.75 | 0.04 | | | | | | | 82.78 | 11.58 | 0.03 |
| 63.94 | 12.75 | 0.04 | | | | | | | 83.02 | 11.56 | 0.03 |
| 64.43 | 12.73 | 0.04 | | | | | | | 83.41 | 11.53 | 0.03 |
| 64.80 | 12.72 | 0.04 | | | | | | | 83.73 | 11.54 | 0.03 |
| 64.53 | 12.69 | 0.04 | | | | | | | 84.15 | 11.54 | 0.03 |
| 65.15 | 12.69 | 0.04 | | | | | | | 84.42 | 11.53 | 0.03 |
| 65.93 | 12.70 | 0.04 | | | | | | | 84.96 | 11.52 | 0.03 |
| 65.74 | 12.68 | 0.04 | | | | | | | 85.29 | 11.51 | 0.03 |
| 65.75 | 12.65 | 0.04 | | | | | | | 89.03 | 11.41 | 0.03 |
| 65.50 | 12.65 | 0.04 | | | | | | | 88.95 | 11.40 | 0.03 |
| 65.63 | 12.63 | 0.04 | | | | | | | 88.92 | 11.39 | 0.03 |
| 65.93 | 12.63 | 0.04 | | | | | | | 89.00 | 11.39 | 0.03 |
| 66.01 | 12.61 | 0.04 | | | | | | | 88.99 | 11.39 | 0.03 |
| 66.55 | 12.62 | 0.04 | | | | | | | 88.93 | 11.38 | 0.03 |
| 66.51 | 12.61 | 0.04 | | | | | | | 89.08 | 11.37 | 0.03 |
| 66.98 | 12.59 | 0.04 | | | | | | | 89.14 | 11.38 | 0.03 |
| 67.21 | 12.57 | 0.04 | | | | | | | 89.39 | 11.37 | 0.03 |
| 67.49 | 12.59 | 0.04 | | | | | | | 89.48 | 11.38 | 0.03 |
| 67.87 | 12.56 | 0.04 | | | | | | | 89.60 | 11.37 | 0.03 |
| 68.26 | 12.57 | 0.04 | | | | | | | 89.73 | 11.36 | 0.03 |
| 68.29 | 12.55 | 0.04 | | | | | | | 90.27 | 11.36 | 0.03 |
| 67.84 | 12.54 | 0.04 | | | | | | | 90.47 | 11.36 | 0.03 |
| 68.05 | 12.51 | 0.04 | | | | | | | 90.69 | 11.33 | 0.03 |
| 69.24 | 12.53 | 0.04 | | | | | | | 91.04 | 11.33 | 0.03 |
| 69.93 | 12.51 | 0.04 | | | | | | | 91.59 | 11.31 | 0.03 |
| 69.40 | 12.50 | 0.04 | | | | | | | 92.35 | 11.30 | 0.03 |
| 68.70 | 12.49 | 0.04 | | | | | | | 93.36 | 11.28 | 0.03 |
| 69.19 | 12.49 | 0.04 | | | | | | | 94.21 | 11.24 | 0.03 |
| 69.41 | 12.47 | 0.04 | | | | | | | 94.93 | 11.24 | 0.03 |
| 69.49 | 12.47 | 0.04 | | | | | | | 95.89 | 11.20 | 0.03 |
| 71.75 | 12.46 | 0.04 | | | | | | | 96.58 | 11.19 | 0.03 |
| 69.91 | 12.44 | 0.04 | | | | | | | 97.46 | 11.17 | 0.03 |
| 71.10 | 12.44 | 0.04 | | | | | | | 98.30 | 11.14 | 0.03 |
| 69.77 | 12.44 | 0.04 | | | | | | | 98.95 | 11.12 | 0.03 |
| 71.74 | 12.43 | 0.04 | | | | | | | 99.85 | 11.11 | 0.03 |
| 71.49 | 12.41 | 0.04 | | | | | | | 100.71 | 11.09 | 0.03 |
| 70.45 | 12.40 | 0.04 | | | | | | | 101.49 | 11.06 | 0.03 |
| 73.14 | 12.41 | 0.04 | | | | | | | 102.24 | 11.06 | 0.03 |
| 72.28 | 12.38 | 0.04 | | | | | | | 103.15 | 11.02 | 0.03 |
| 72.00 | 12.39 | 0.04 | | | | | | | 103.80 | 10.99 | 0.03 |
| 72.81 | 12.37 | 0.04 | | | | | | | 104.52 | 10.97 | 0.03 |
| 73.34 | 12.35 | 0.04 | | | | | | | 105.65 | 10.94 | 0.03 |
| 73.49 | 12.35 | 0.04 | | | | | | | 106.50 | 10.95 | 0.03 |
| 72.24 | 12.34 | 0.04 | | | | | | | 107.02 | 10.92 | 0.03 |
| 74.17 | 12.34 | 0.04 | | | | | | | 108.08 | 10.89 | 0.03 |
| 73.63 | 12.34 | 0.04 | | | | | | | 108.66 | 10.86 | 0.03 |
| 73.44 | 12.31 | 0.04 | | | | | | | 109.48 | 10.85 | 0.03 |
| 74.35 | 12.30 | 0.04 | | | | | | | 110.26 | 10.83 | 0.03 |
| 74.81 | 12.30 | 0.04 | | | | | | | 110.84 | 10.82 | 0.03 |
| 74.66 | 12.31 | 0.04 | | | | | | | 111.45 | 10.81 | 0.03 |
| 75.11 | 12.28 | 0.04 | | | | | | | 112.51 | 10.79 | 0.03 |
| 74.65 | 12.30 | 0.04 | | | | | | | 113.12 | 10.77 | 0.03 |
| 74.10 | 12.27 | 0.04 | | | | | | | 113.81 | 10.76 | 0.03 |
| 76.47 | 12.28 | 0.04 | | | | | | | 114.38 | 10.74 | 0.03 |
| 75.14 | 12.27 | 0.04 | | | | | | | 115.56 | 10.71 | 0.03 |
| 74.98 | 12.25 | 0.04 | | | | | | | 115.93 | 10.71 | 0.03 |
| 75.56 | 12.24 | 0.04 | | | | | | | 116.70 | 10.69 | 0.03 |
| 76.66 | 12.24 | 0.04 | | | | | | | 117.67 | 10.67 | 0.03 |
| 76.47 | 12.24 | 0.04 | | | | | | | 118.28 | 10.66 | 0.03 |
| 76.55 | 12.21 | 0.04 | | | | | | | 119.16 | 10.64 | 0.03 |
| 77.47 | 12.22 | 0.04 | | | | | | | 119.43 | 10.62 | 0.03 |
| 76.98 | 12.20 | 0.04 | | | | | | | 120.30 | 10.61 | 0.03 |
| 77.49 | 12.21 | 0.04 | | | | | | | 121.03 | 10.60 | 0.03 |
| 78.24 | 12.20 | 0.04 | | | | | | | 122.51 | 10.56 | 0.03 |
| 77.41 | 12.18 | 0.04 | | | | | | | 125.30 | 10.52 | 0.03 |
| 77.53 | 12.19 | 0.04 | | | | | | | 128.08 | 10.46 | 0.03 |
| 78.09 | 12.17 | 0.04 | | | | | | | 130.70 | 10.40 | 0.03 |
| 78.38 | 12.18 | 0.04 | | | | | | | 133.22 | 10.34 | 0.03 |
| 78.26 | 12.17 | 0.04 | | | | | | | 135.69 | 10.31 | 0.03 |
| 77.82 | 12.15 | 0.04 | | | | | | | 138.13 | 10.26 | 0.03 |
| 78.91 | 12.15 | 0.04 | | | | | | | 140.45 | 10.21 | 0.03 |
| 78.65 | 12.14 | 0.04 | | | | | | | 146.36 | 10.12 | 0.03 |
| 79.04 | 12.11 | 0.04 | | | | | | | | | |
| 78.49 | 12.14 | 0.04 | | | | | | | | | |
| 79.74 | 12.11 | 0.04 | | | | | | | | | |
| 79.34 | 12.10 | 0.04 | | | | | | | | | |
| 80.44 | 12.10 | 0.04 | | | | | | | | | |
| 79.79 | 12.10 | 0.04 | | | | | | | | | |
| 80.38 | 12.10 | 0.04 | | | | | | | | | |
| 79.57 | 12.07 | 0.04 | | | | | | | | | |
| 80.81 | 12.07 | 0.04 | | | | | | | | | |
| 79.66 | 12.06 | 0.04 | | | | | | | | | |
| 81.10 | 12.07 | 0.04 | | | | | | | | | |
| 81.42 | 12.07 | 0.04 | | | | | | | | | |
| 79.89 | 12.03 | 0.04 | | | | | | | | | |
| 80.53 | 12.04 | 0.04 | | | | | | | | | |
| 81.13 | 12.03 | 0.04 | | | | | | | | | |
| 81.06 | 12.04 | 0.04 | | | | | | | | | |
| 80.88 | 12.02 | 0.04 | | | | | | | | | |
| 82.38 | 12.01 | 0.04 | | | | | | | | | |
| 82.66 | 11.99 | 0.04 | | | | | | | | | |
| 81.90 | 11.99 | 0.04 | | | | | | | | | |
| 81.98 | 11.96 | 0.04 | | | | | | | | | |
| 81.46 | 12.00 | 0.04 | | | | | | | | | |
| 82.14 | 11.97 | 0.04 | | | | | | | | | |
| 82.91 | 11.94 | 0.04 | | | | | | | | | |
| 82.38 | 11.95 | 0.04 | | | | | | | | | |
| 82.46 | 11.97 | 0.04 | | | | | | | | | |
| 81.82 | 11.92 | 0.04 | | | | | | | | | |
| 83.53 | 11.95 | 0.04 | | | | | | | | | |
| 82.59 | 11.95 | 0.04 | | | | | | | | | |
| 83.61 | 11.92 | 0.04 | | | | | | | | | |
| 85.02 | 11.91 | 0.04 | | | | | | | | | |
| 83.50 | 11.92 | 0.04 | | | | | | | | | |
| 85.17 | 11.88 | 0.04 | | | | | | | | | |
| 83.59 | 11.91 | 0.04 | | | | | | | | | |
| 83.90 | 11.87 | 0.04 | | | | | | | | | |
| 84.58 | 11.87 | 0.04 | | | | | | | | | |
| 84.03 | 11.86 | 0.04 | | | | | | | | | |
| 85.18 | 11.88 | 0.04 | | | | | | | | | |
| 84.62 | 11.86 | 0.04 | | | | | | | | | |
| 84.75 | 11.85 | 0.04 | | | | | | | | | |
| 85.35 | 11.82 | 0.04 | | | | | | | | | |
| 85.72 | 11.82 | 0.04 | | | | | | | | | |
| 85.86 | 11.83 | 0.04 | | | | | | | | | |
| 85.55 | 11.82 | 0.04 | | | | | | | | | |
| 84.45 | 11.81 | 0.04 | | | | | | | | | |
| 85.71 | 11.80 | 0.04 | | | | | | | | | |
| 86.51 | 11.81 | 0.04 | | | | | | | | | |
| 85.60 | 11.80 | 0.04 | | | | | | | | | |
| 85.75 | 11.81 | 0.04 | | | | | | | | | |
| 87.25 | 11.79 | 0.04 | | | | | | | | | |
| 87.69 | 11.77 | 0.04 | | | | | | | | | |
| 86.99 | 11.79 | 0.04 | | | | | | | | | |
| 87.43 | 11.77 | 0.04 | | | | | | | | | |
| 86.90 | 11.77 | 0.04 | | | | | | | | | |
| 86.28 | 11.75 | 0.04 | | | | | | | | | |
| 88.38 | 11.75 | 0.04 | | | | | | | | | |
| 88.24 | 11.73 | 0.04 | | | | | | | | | |
| 87.76 | 11.71 | 0.04 | | | | | | | | | |
| 87.66 | 11.73 | 0.04 | | | | | | | | | |
| 87.58 | 11.71 | 0.04 | | | | | | | | | |
| 87.60 | 11.73 | 0.04 | | | | | | | | | |
| 88.15 | 11.70 | 0.04 | | | | | | | | | |
| 88.36 | 11.73 | 0.04 | | | | | | | | | |
| 88.30 | 11.69 | 0.04 | | | | | | | | | |
| 89.05 | 11.71 | 0.04 | | | | | | | | | |
| 87.93 | 11.68 | 0.04 | | | | | | | | | |
| 89.78 | 11.69 | 0.04 | | | | | | | | | |
| 88.59 | 11.68 | 0.04 | | | | | | | | | |
| 88.59 | 11.69 | 0.04 | | | | | | | | | |
| 89.53 | 11.67 | 0.04 | | | | | | | | | |
| 89.92 | 11.67 | 0.04 | | | | | | | | | |
| 90.42 | 11.63 | 0.04 | | | | | | | | | |
| 90.06 | 11.63 | 0.04 | | | | | | | | | |
| 90.22 | 11.61 | 0.04 | | | | | | | | | |
| 90.64 | 11.65 | 0.04 | | | | | | | | | |
| 88.79 | 11.64 | 0.04 | | | | | | | | | |
| 90.36 | 11.62 | 0.04 | | | | | | | | | |
| 90.27 | 11.60 | 0.04 | | | | | | | | | |
| 89.55 | 11.63 | 0.04 | | | | | | | | | |
| 89.69 | 11.59 | 0.04 | | | | | | | | | |
| 92.41 | 11.62 | 0.04 | | | | | | | | | |
| 92.42 | 11.61 | 0.04 | | | | | | | | | |
| 90.79 | 11.59 | 0.04 | | | | | | | | | |
| 92.96 | 11.59 | 0.04 | | | | | | | | | |
| 90.70 | 11.60 | 0.04 | | | | | | | | | |
| 92.40 | 11.56 | 0.04 | | | | | | | | | |
| 91.22 | 11.57 | 0.04 | | | | | | | | | |

**Table S2:** Calculated pressure-volume (P-V) data of $H_2O$ ice phases recorded in four different compression runs presented in Figure 6.

| **$H_2O$ ice coexisting with AHH-D(I)MA** | | | | | | **$H_2O$ ice coexisting with DMA’** | | | | | |
|---|---|---|---|---|---|---|---|---|---|---|---|
| **dDAC12** | | | **mDAC1** | | | **dDAC13** | | | **mDAC2** | | |
| **P (GPa)** | **V (Å$^3$)** | **Err (Å)** | **P (GPa)** | **V (Å$^3$)** | **Err (Å)** | **P (GPa)** | **V (Å$^3$)** | **Err (Å)** | **P (GPa)** | **V (Å$^3$)** | **Err (Å)** |
| 6.03 | 16.80 | 0.05 | 6.20 | 16.88 | 0.05 | 8.93 | 15.82 | 0.05 | 8.30 | 16.32 | 0.05 |
| 6.13 | 16.79 | 0.05 | 8.10 | 16.28 | 0.05 | 9.18 | 15.73 | 0.05 | 8.30 | 16.27 | 0.05 |
| 6.15 | 16.78 | 0.05 | 9.20 | 15.91 | 0.05 | 9.54 | 15.63 | 0.05 | 8.40 | 16.24 | 0.05 |
| 6.12 | 16.77 | 0.05 | 12.20 | 15.17 | 0.05 | 17.10 | 14.05 | 0.05 | 8.50 | 16.20 | 0.05 |
| 6.15 | 16.76 | 0.05 | 13.00 | 15.01 | 0.05 | 17.89 | 13.92 | 0.05 | 8.70 | 16.16 | 0.05 |
| 6.12 | 16.74 | 0.05 | 13.90 | 14.84 | 0.05 | 18.54 | 13.83 | 0.05 | 8.90 | 16.10 | 0.05 |
| 6.24 | 16.73 | 0.05 | 15.40 | 14.67 | 0.05 | 19.27 | 13.71 | 0.05 | 9.20 | 16.04 | 0.05 |
| 6.26 | 16.72 | 0.05 | 16.80 | 14.54 | 0.05 | 19.60 | 13.67 | 0.05 | 9.40 | 16.01 | 0.05 |
| 6.16 | 16.70 | 0.05 | 17.50 | 14.36 | 0.05 | 19.89 | 13.63 | 0.05 | 9.50 | 15.94 | 0.05 |
| 6.26 | 16.69 | 0.05 | 18.00 | 14.23 | 0.05 | 20.48 | 13.54 | 0.05 | 9.60 | 15.90 | 0.05 |
| 6.27 | 16.67 | 0.05 | 19.50 | 14.16 | 0.05 | 20.77 | 13.51 | 0.05 | 9.90 | 15.86 | 0.05 |
| 6.34 | 16.66 | 0.05 | 20.30 | 14.02 | 0.05 | 21.12 | 13.46 | 0.05 | 10.10 | 15.82 | 0.05 |
| 6.39 | 16.64 | 0.05 | 21.10 | 13.89 | 0.05 | 21.57 | 13.41 | 0.05 | 10.20 | 15.79 | 0.05 |
| 6.54 | 16.62 | 0.05 | 22.00 | 13.76 | 0.05 | 22.04 | 13.35 | 0.05 | 10.20 | 15.74 | 0.05 |
| 6.54 | 16.61 | 0.05 | 23.30 | 13.65 | 0.05 | 22.40 | 13.30 | 0.05 | 10.40 | 15.69 | 0.05 |
| 6.77 | 16.58 | 0.05 | 24.20 | 13.52 | 0.05 | 22.55 | 13.27 | 0.05 | 10.50 | 15.63 | 0.05 |
| 6.75 | 16.56 | 0.05 | 25.60 | 13.38 | 0.05 | 22.99 | 13.23 | 0.05 | 10.70 | 15.60 | 0.05 |
| 6.86 | 16.54 | 0.05 | 26.50 | 13.31 | 0.05 | 23.36 | 13.19 | 0.05 | 10.90 | 15.54 | 0.05 |
| 6.76 | 16.52 | 0.05 | 27.30 | 13.25 | 0.05 | 23.69 | 13.15 | 0.05 | 11.10 | 15.48 | 0.05 |
| 6.92 | 16.50 | 0.05 | 28.10 | 13.07 | 0.05 | 23.88 | 13.14 | 0.05 | 11.30 | 15.40 | 0.05 |
| 6.99 | 16.48 | 0.05 | 28.80 | 13.03 | 0.05 | 24.28 | 13.09 | 0.05 | 11.60 | 15.36 | 0.05 |
| 7.06 | 16.46 | 0.05 | 30.20 | 12.93 | 0.05 | 24.70 | 13.05 | 0.05 | 11.80 | 15.29 | 0.05 |
| 7.11 | 16.44 | 0.05 | 31.70 | 12.82 | 0.05 | 25.16 | 13.00 | 0.05 | 12.10 | 15.25 | 0.05 |
| 7.35 | 16.41 | 0.05 | 32.40 | 12.73 | 0.05 | 25.65 | 12.94 | 0.05 | 12.50 | 15.18 | 0.05 |
| 7.21 | 16.39 | 0.05 | 33.33 | 12.57 | 0.05 | 26.18 | 12.88 | 0.05 | 12.90 | 15.11 | 0.05 |
| 7.34 | 16.37 | 0.05 | 33.51 | 12.57 | 0.05 | 26.78 | 12.82 | 0.05 | 13.20 | 15.08 | 0.05 |
| 7.41 | 16.35 | 0.05 | 33.62 | 12.56 | 0.05 | 27.21 | 12.77 | 0.05 | 13.50 | 15.02 | 0.05 |
| 7.49 | 16.33 | 0.05 | 33.87 | 12.53 | 0.05 | 27.66 | 12.72 | 0.05 | 13.80 | 14.94 | 0.05 |
| 7.58 | 16.31 | 0.05 | 34.82 | 12.50 | 0.05 | 28.25 | 12.67 | 0.05 | 14.10 | 14.89 | 0.05 |
| 7.80 | 16.28 | 0.05 | 35.33 | 12.45 | 0.05 | 28.80 | 12.61 | 0.05 | 14.40 | 14.87 | 0.05 |
| 7.68 | 16.26 | 0.05 | 38.62 | 12.10 | 0.05 | 29.31 | 12.56 | 0.05 | 14.50 | 14.81 | 0.05 |
| 7.97 | 16.24 | 0.05 | 38.91 | 12.10 | 0.05 | 29.85 | 12.51 | 0.05 | 14.70 | 14.77 | 0.05 |
| 8.03 | 16.22 | 0.05 | 40.36 | 12.02 | 0.05 | 30.28 | 12.47 | 0.05 | 14.90 | 14.72 | 0.05 |
| 8.12 | 16.20 | 0.05 | 41.30 | 11.96 | 0.05 | 31.22 | 12.40 | 0.05 | 15.20 | 14.67 | 0.05 |
| 8.20 | 16.18 | 0.05 | 42.11 | 11.89 | 0.05 | 31.69 | 12.35 | 0.05 | 15.40 | 14.61 | 0.05 |
| 8.27 | 16.15 | 0.05 | 42.89 | 11.84 | 0.05 | 32.34 | 12.31 | 0.05 | 15.60 | 14.56 | 0.05 |
| 8.43 | 16.13 | 0.05 | 43.92 | 11.79 | 0.05 | 32.80 | 12.27 | 0.05 | 15.80 | 14.50 | 0.05 |
| 8.42 | 16.10 | 0.05 | 45.78 | 11.67 | 0.05 | 33.31 | 12.22 | 0.05 | 16.00 | 14.44 | 0.05 |
| 8.70 | 16.08 | 0.05 | 47.08 | 11.61 | 0.05 | 33.87 | 12.18 | 0.05 | 16.30 | 14.40 | 0.05 |
| 8.71 | 16.05 | 0.05 | 47.91 | 11.59 | 0.05 | 34.50 | 12.13 | 0.05 | 16.40 | 14.34 | 0.05 |
| 9.06 | 16.03 | 0.05 | 48.35 | 11.57 | 0.05 | 35.08 | 12.08 | 0.05 | 16.60 | 14.28 | 0.05 |
| 8.97 | 16.00 | 0.05 | 49.25 | 11.50 | 0.05 | 35.63 | 12.04 | 0.05 | 16.80 | 14.23 | 0.05 |
| 9.11 | 15.98 | 0.05 | 49.60 | 11.49 | 0.05 | 36.22 | 11.99 | 0.05 | 17.00 | 14.16 | 0.05 |
| 9.25 | 15.95 | 0.05 | 50.28 | 11.46 | 0.05 | 36.79 | 11.95 | 0.05 | 17.20 | 14.11 | 0.05 |
| 9.23 | 15.93 | 0.05 | 50.50 | 11.43 | 0.05 | 37.33 | 11.92 | 0.05 | 17.50 | 14.03 | 0.05 |
| 9.18 | 15.90 | 0.05 | 50.97 | 11.41 | 0.05 | 37.89 | 11.87 | 0.05 | 17.70 | 13.97 | 0.05 |
| 9.40 | 15.88 | 0.05 | 51.28 | 11.38 | 0.05 | 38.48 | 11.83 | 0.05 | 17.90 | 13.91 | 0.05 |
| 9.35 | 15.85 | 0.05 | 51.73 | 11.36 | 0.05 | 39.32 | 11.78 | 0.05 | 18.10 | 13.84 | 0.05 |
| 9.64 | 15.82 | 0.05 | 52.53 | 11.34 | 0.05 | 40.05 | 11.74 | 0.05 | 18.40 | 13.81 | 0.05 |
| 9.69 | 15.79 | 0.05 | 53.22 | 11.28 | 0.05 | 40.49 | 11.70 | 0.05 | 18.60 | 13.76 | 0.05 |
| 9.70 | 15.77 | 0.05 | 54.46 | 11.23 | 0.05 | 41.05 | 11.67 | 0.05 | 18.80 | 13.72 | 0.05 |
| 9.86 | 15.74 | 0.05 | 55.01 | 11.19 | 0.05 | 41.55 | 11.63 | 0.05 | 19.00 | 13.67 | 0.05 |
| 10.11 | 15.71 | 0.05 | 55.31 | 11.17 | 0.05 | 42.40 | 11.58 | 0.05 | 19.10 | 13.64 | 0.05 |
| 10.08 | 15.68 | 0.05 | 57.23 | 11.12 | 0.05 | 42.95 | 11.55 | 0.05 | 19.30 | 13.61 | 0.05 |
| 10.24 | 15.65 | 0.05 | 57.69 | 11.10 | 0.05 | 43.49 | 11.51 | 0.05 | 19.40 | 13.60 | 0.05 |
| 10.33 | 15.62 | 0.05 | 58.47 | 11.06 | 0.05 | 44.12 | 11.47 | 0.05 | 19.70 | 13.56 | 0.05 |
| 10.38 | 15.59 | 0.05 | 58.95 | 11.03 | 0.05 | 44.65 | 11.44 | 0.05 | 19.90 | 13.54 | 0.05 |
| 10.62 | 15.56 | 0.05 | 59.58 | 10.99 | 0.05 | 45.31 | 11.39 | 0.05 | 20.10 | 13.51 | 0.05 |
| 10.73 | 15.53 | 0.05 | 60.10 | 10.98 | 0.05 | 45.84 | 11.36 | 0.05 | 20.30 | 13.50 | 0.05 |
| 10.62 | 15.50 | 0.05 | 60.71 | 10.95 | 0.05 | 46.41 | 11.32 | 0.05 | 20.50 | 13.44 | 0.05 |
| 10.91 | 15.46 | 0.05 | 61.67 | 10.91 | 0.05 | 46.98 | 11.28 | 0.05 | 20.60 | 13.43 | 0.05 |
| 10.68 | 15.43 | 0.05 | 64.19 | 10.82 | 0.05 | 47.59 | 11.25 | 0.05 | 20.70 | 13.39 | 0.05 |
| 10.93 | 15.40 | 0.05 | 65.38 | 10.76 | 0.05 | 48.09 | 11.21 | 0.05 | 20.90 | 13.37 | 0.05 |
| 10.87 | 15.37 | 0.05 | 66.52 | 10.71 | 0.05 | 48.79 | 11.17 | 0.05 | 21.00 | 13.32 | 0.05 |
| 11.10 | 15.34 | 0.05 | 70.40 | 10.56 | 0.05 | 49.37 | 11.13 | 0.05 | 21.63 | 13.29 | 0.05 |
| 11.15 | 15.31 | 0.05 | 71.14 | 10.54 | 0.05 | 49.93 | 11.10 | 0.05 | 21.86 | 13.25 | 0.05 |
| 11.36 | 15.28 | 0.05 | | | | 50.37 | 11.07 | 0.05 | 22.10 | 13.23 | 0.05 |
| 11.64 | 15.25 | 0.05 | | | | 50.55 | 11.05 | 0.05 | 22.31 | 13.20 | 0.05 |
| 11.67 | 15.21 | 0.05 | | | | 50.86 | 11.03 | 0.05 | 22.49 | 13.19 | 0.05 |
| 11.84 | 15.18 | 0.05 | | | | 51.21 | 11.01 | 0.05 | 22.72 | 13.16 | 0.05 |
| 11.86 | 15.15 | 0.05 | | | | 51.43 | 10.99 | 0.05 | 22.93 | 13.15 | 0.05 |
| 11.98 | 15.11 | 0.05 | | | | 51.69 | 10.98 | 0.05 | 23.13 | 13.12 | 0.05 |
| 12.12 | 15.08 | 0.05 | | | | 52.06 | 10.96 | 0.05 | 23.36 | 13.11 | 0.05 |
| 12.10 | 15.04 | 0.05 | | | | 52.12 | 10.95 | 0.05 | 23.58 | 13.08 | 0.05 |
| 12.13 | 15.00 | 0.05 | | | | 52.38 | 10.94 | 0.05 | 23.78 | 13.05 | 0.05 |
| 12.12 | 14.97 | 0.05 | | | | 52.78 | 10.92 | 0.05 | 23.94 | 13.01 | 0.05 |
| 12.75 | 14.93 | 0.05 | | | | 53.07 | 10.90 | 0.05 | 24.15 | 12.98 | 0.05 |
| 12.75 | 14.90 | 0.05 | | | | 53.65 | 10.88 | 0.05 | 24.35 | 12.93 | 0.05 |
| 13.01 | 14.86 | 0.05 | | | | 53.93 | 10.86 | 0.05 | 24.58 | 12.92 | 0.05 |
| 13.15 | 14.83 | 0.05 | | | | 54.59 | 10.83 | 0.05 | 24.82 | 12.88 | 0.05 |
| 13.06 | 14.79 | 0.05 | | | | 54.98 | 10.80 | 0.05 | 25.04 | 12.85 | 0.05 |
| 13.08 | 14.75 | 0.05 | | | | 55.52 | 10.79 | 0.05 | 25.28 | 12.81 | 0.05 |
| 13.26 | 14.72 | 0.05 | | | | 56.08 | 10.76 | 0.05 | 25.56 | 12.79 | 0.05 |
| 13.56 | 14.68 | 0.05 | | | | 56.48 | 10.75 | 0.05 | 25.82 | 12.74 | 0.05 |
| 14.10 | 14.64 | 0.05 | | | | 56.92 | 10.73 | 0.05 | 26.09 | 12.72 | 0.05 |
| 15.10 | 14.38 | 0.05 | | | | 57.50 | 10.71 | 0.05 | 26.32 | 12.69 | 0.05 |
| 15.41 | 14.35 | 0.05 | | | | 58.10 | 10.68 | 0.05 | 26.59 | 12.66 | 0.05 |
| 15.66 | 14.31 | 0.05 | | | | 58.42 | 10.66 | 0.05 | 26.84 | 12.64 | 0.05 |
| 16.00 | 14.27 | 0.05 | | | | 59.00 | 10.65 | 0.05 | 27.08 | 12.61 | 0.05 |
| 16.38 | 14.23 | 0.05 | | | | 59.41 | 10.64 | 0.05 | 27.32 | 12.58 | 0.05 |
| 16.69 | 14.19 | 0.05 | | | | 60.00 | 10.61 | 0.05 | 27.54 | 12.55 | 0.05 |
| 17.07 | 14.15 | 0.05 | | | | 60.51 | 10.60 | 0.05 | 27.77 | 12.52 | 0.05 |
| 17.25 | 14.11 | 0.05 | | | | 60.93 | 10.59 | 0.05 | 28.02 | 12.51 | 0.05 |
| 17.43 | 14.07 | 0.05 | | | | 61.68 | 10.56 | 0.05 | 28.25 | 12.50 | 0.05 |
| 17.71 | 14.03 | 0.05 | | | | 62.17 | 10.55 | 0.05 | 28.56 | 12.47 | 0.05 |
| 17.74 | 13.99 | 0.05 | | | | 62.64 | 10.53 | 0.05 | 28.83 | 12.44 | 0.05 |
| 18.09 | 13.96 | 0.05 | | | | 62.99 | 10.52 | 0.05 | 29.11 | 12.42 | 0.05 |
| 18.54 | 13.92 | 0.05 | | | | 63.61 | 10.50 | 0.05 | 29.37 | 12.37 | 0.05 |
| 18.77 | 13.88 | 0.05 | | | | 64.01 | 10.48 | 0.05 | 29.65 | 12.35 | 0.05 |
| 19.03 | 13.84 | 0.05 | | | | 64.46 | 10.48 | 0.05 | 29.97 | 12.33 | 0.05 |
| 19.41 | 13.80 | 0.05 | | | | 64.88 | 10.46 | 0.05 | 30.24 | 12.32 | 0.05 |
| 19.53 | 13.76 | 0.05 | | | | 65.43 | 10.44 | 0.05 | 30.49 | 12.29 | 0.05 |
| 19.67 | 13.72 | 0.05 | | | | 66.03 | 10.43 | 0.05 | 30.84 | 12.25 | 0.05 |
| 19.84 | 13.69 | 0.05 | | | | 66.34 | 10.42 | 0.05 | 31.17 | 12.23 | 0.05 |
| 19.95 | 13.65 | 0.05 | | | | 66.78 | 10.41 | 0.05 | 31.45 | 12.22 | 0.05 |
| 20.25 | 13.61 | 0.05 | | | | 67.14 | 10.40 | 0.05 | 31.76 | 12.19 | 0.05 |
| 20.47 | 13.57 | 0.05 | | | | 67.65 | 10.39 | 0.05 | 32.05 | 12.16 | 0.05 |
| 20.56 | 13.54 | 0.05 | | | | 68.00 | 10.38 | 0.05 | 32.34 | 12.13 | 0.05 |
| 20.96 | 13.50 | 0.05 | | | | 68.44 | 10.37 | 0.05 | 32.60 | 12.10 | 0.05 |
| 21.07 | 13.47 | 0.05 | | | | 68.90 | 10.36 | 0.05 | 32.94 | 12.08 | 0.05 |
| 21.49 | 13.43 | 0.05 | | | | 69.31 | 10.34 | 0.05 | 33.22 | 12.07 | 0.05 |
| 21.87 | 13.40 | 0.05 | | | | 69.69 | 10.34 | 0.05 | 33.57 | 12.01 | 0.05 |
| 22.24 | 13.37 | 0.05 | | | | 70.08 | 10.33 | 0.05 | 33.89 | 11.99 | 0.05 |
| 22.45 | 13.34 | 0.05 | | | | 70.54 | 10.31 | 0.05 | 34.16 | 11.96 | 0.05 |
| 22.66 | 13.31 | 0.05 | | | | 70.98 | 10.30 | 0.05 | 34.46 | 11.95 | 0.05 |
| 23.06 | 13.28 | 0.05 | | | | 71.38 | 10.30 | 0.05 | 34.79 | 11.94 | 0.05 |
| 23.46 | 13.26 | 0.05 | | | | 71.91 | 10.29 | 0.05 | 35.13 | 11.92 | 0.05 |
| 28.52 | 12.83 | 0.05 | | | | 72.17 | 10.28 | 0.05 | 35.43 | 11.92 | 0.05 |
| 28.49 | 12.81 | 0.05 | | | | 72.84 | 10.26 | 0.05 | 35.75 | 11.91 | 0.05 |
| 28.92 | 12.79 | 0.05 | | | | 73.32 | 10.25 | 0.05 | 36.07 | 11.90 | 0.05 |
| 29.29 | 12.78 | 0.05 | | | | 73.82 | 10.24 | 0.05 | 36.39 | 11.90 | 0.05 |
| 28.93 | 12.77 | 0.05 | | | | 74.22 | 10.23 | 0.05 | 36.65 | 11.88 | 0.05 |
| 28.99 | 12.76 | 0.05 | | | | 74.63 | 10.22 | 0.05 | | | |
| 29.67 | 12.74 | 0.05 | | | | 74.97 | 10.21 | 0.05 | | | |
| 30.05 | 12.73 | 0.05 | | | | 75.30 | 10.20 | 0.05 | | | |
| 30.19 | 12.72 | 0.05 | | | | 75.81 | 10.19 | 0.05 | | | |
| 30.74 | 12.70 | 0.05 | | | | 76.20 | 10.18 | 0.05 | | | |
| 30.42 | 12.69 | 0.05 | | | | 76.50 | 10.18 | 0.05 | | | |
| 30.79 | 12.68 | 0.05 | | | | 77.06 | 10.17 | 0.05 | | | |
| 31.36 | 12.67 | 0.05 | | | | 77.23 | 10.16 | 0.05 | | | |
| 31.82 | 12.65 | 0.05 | | | | 77.66 | 10.15 | 0.05 | | | |
| 32.19 | 12.64 | 0.05 | | | | 77.81 | 10.14 | 0.05 | | | |
| 32.34 | 12.63 | 0.05 | | | | 78.18 | 10.12 | 0.05 | | | |
| 33.00 | 12.61 | 0.05 | | | | 78.62 | 10.12 | 0.05 | | | |
| 32.71 | 12.59 | 0.05 | | | | 79.06 | 10.12 | 0.05 | | | |
| 33.02 | 12.58 | 0.05 | | | | 79.44 | 10.11 | 0.05 | | | |
| 34.21 | 12.56 | 0.05 | | | | 79.70 | 10.10 | 0.05 | | | |
| 34.26 | 12.55 | 0.05 | | | | 80.19 | 10.09 | 0.05 | | | |
| 34.47 | 12.53 | 0.05 | | | | 80.53 | 10.08 | 0.05 | | | |
| 34.14 | 12.52 | 0.05 | | | | 81.00 | 10.07 | 0.05 | | | |
| 34.63 | 12.50 | 0.05 | | | | 81.63 | 10.06 | 0.05 | | | |
| 34.78 | 12.48 | 0.05 | | | | 81.96 | 10.05 | 0.05 | | | |
| 34.85 | 12.47 | 0.05 | | | | 82.31 | 10.04 | 0.05 | | | |
| 35.12 | 12.45 | 0.05 | | | | 82.78 | 10.03 | 0.05 | | | |
| 35.44 | 12.43 | 0.05 | | | | 83.02 | 10.03 | 0.05 | | | |
| 35.65 | 12.42 | 0.05 | | | | 83.41 | 10.02 | 0.05 | | | |
| 35.79 | 12.40 | 0.05 | | | | 83.73 | 10.01 | 0.05 | | | |
| 36.22 | 12.39 | 0.05 | | | | 84.15 | 10.01 | 0.05 | | | |
| 36.93 | 12.37 | 0.05 | | | | 84.42 | 10.00 | 0.05 | | | |
| 36.99 | 12.36 | 0.05 | | | | 84.96 | 9.99 | 0.05 | | | |
| 36.89 | 12.35 | 0.05 | | | | 85.29 | 9.98 | 0.05 | | | |
| 37.10 | 12.33 | 0.05 | | | | 89.03 | 9.90 | 0.05 | | | |
| 37.19 | 12.31 | 0.05 | | | | 88.95 | 9.90 | 0.05 | | | |
| 37.78 | 12.31 | 0.05 | | | | 88.92 | 9.90 | 0.05 | | | |
| 37.93 | 12.29 | 0.05 | | | | 89.00 | 9.90 | 0.05 | | | |
| 37.45 | 12.27 | 0.05 | | | | 88.99 | 9.90 | 0.05 | | | |
| 37.92 | 12.26 | 0.05 | | | | 88.93 | 9.90 | 0.05 | | | |
| 38.23 | 12.24 | 0.05 | | | | 89.08 | 9.90 | 0.05 | | | |
| 38.37 | 12.23 | 0.05 | | | | 89.14 | 9.90 | 0.05 | | | |
| 38.64 | 12.21 | 0.05 | | | | 89.39 | 9.90 | 0.05 | | | |
| 38.72 | 12.20 | 0.05 | | | | 89.48 | 9.90 | 0.05 | | | |
| 39.15 | 12.17 | 0.05 | | | | 89.60 | 9.90 | 0.05 | | | |
| 39.69 | 12.15 | 0.05 | | | | 89.73 | 9.90 | 0.05 | | | |
| 40.02 | 12.14 | 0.05 | | | | 90.27 | 9.88 | 0.05 | | | |
| 40.30 | 12.12 | 0.05 | | | | 90.47 | 9.88 | 0.05 | | | |
| 40.81 | 12.09 | 0.05 | | | | 90.69 | 9.87 | 0.05 | | | |
| 40.68 | 12.07 | 0.05 | | | | 91.04 | 9.87 | 0.05 | | | |
| 41.41 | 12.06 | 0.05 | | | | 91.59 | 9.85 | 0.05 | | | |
| 41.54 | 12.03 | 0.05 | | | | 92.35 | 9.84 | 0.05 | | | |
| 41.59 | 12.02 | 0.05 | | | | 93.36 | 9.83 | 0.05 | | | |
| 42.06 | 11.99 | 0.05 | | | | 94.21 | 9.81 | 0.05 | | | |
| 42.27 | 11.96 | 0.05 | | | | 94.93 | 9.79 | 0.05 | | | |
| 42.78 | 11.95 | 0.05 | | | | 95.89 | 9.77 | 0.05 | | | |
| 42.59 | 11.91 | 0.05 | | | | 96.58 | 9.76 | 0.05 | | | |
| 42.82 | 11.88 | 0.05 | | | | 97.46 | 9.75 | 0.05 | | | |
| 43.06 | 11.87 | 0.05 | | | | 98.30 | 9.73 | 0.05 | | | |
| 43.28 | 11.83 | 0.05 | | | | 98.95 | 9.71 | 0.05 | | | |
| 43.60 | 11.81 | 0.05 | | | | 99.85 | 9.70 | 0.05 | | | |
| 44.12 | 11.79 | 0.05 | | | | 100.71 | 9.68 | 0.05 | | | |
| 43.29 | 11.76 | 0.05 | | | | 101.49 | 9.67 | 0.05 | | | |
| 44.19 | 11.73 | 0.05 | | | | 102.24 | 9.66 | 0.05 | | | |
| 44.90 | 11.72 | 0.05 | | | | 103.15 | 9.64 | 0.05 | | | |
| 44.83 | 11.69 | 0.05 | | | | 103.80 | 9.62 | 0.05 | | | |
| 44.99 | 11.68 | 0.05 | | | | 104.52 | 9.61 | 0.05 | | | |
| 45.22 | 11.65 | 0.05 | | | | 105.65 | 9.59 | 0.05 | | | |
| 45.20 | 11.63 | 0.05 | | | | 106.50 | 9.58 | 0.05 | | | |
| 45.56 | 11.62 | 0.05 | | | | 107.02 | 9.57 | 0.05 | | | |
| 45.66 | 11.60 | 0.05 | | | | 108.08 | 9.55 | 0.05 | | | |
| 46.75 | 11.57 | 0.05 | | | | 108.66 | 9.55 | 0.05 | | | |
| 46.25 | 11.56 | 0.05 | | | | 109.48 | 9.52 | 0.05 | | | |
| 46.29 | 11.54 | 0.05 | | | | 110.26 | 9.51 | 0.05 | | | |
| 46.37 | 11.52 | 0.05 | | | | 110.84 | 9.50 | 0.05 | | | |
| 47.27 | 11.49 | 0.05 | | | | 111.45 | 9.49 | 0.05 | | | |
| 47.16 | 11.48 | 0.05 | | | | 112.51 | 9.48 | 0.05 | | | |
| 47.17 | 11.47 | 0.05 | | | | 113.12 | 9.46 | 0.05 | | | |
| 47.44 | 11.46 | 0.05 | | | | 113.81 | 9.45 | 0.05 | | | |
| 47.38 | 11.44 | 0.05 | | | | 114.38 | 9.44 | 0.05 | | | |
| 48.27 | 11.42 | 0.05 | | | | 115.56 | 9.42 | 0.05 | | | |
| 48.13 | 11.41 | 0.05 | | | | 115.93 | 9.41 | 0.05 | | | |
| 48.64 | 11.39 | 0.05 | | | | 116.70 | 9.40 | 0.05 | | | |
| 48.99 | 11.38 | 0.05 | | | | 117.67 | 9.39 | 0.05 | | | |
| 49.35 | 11.37 | 0.05 | | | | 118.28 | 9.38 | 0.05 | | | |
| 49.45 | 11.35 | 0.05 | | | | 119.16 | 9.37 | 0.05 | | | |
| 49.19 | 11.35 | 0.05 | | | | 119.43 | 9.35 | 0.05 | | | |
| 49.88 | 11.33 | 0.05 | | | | 120.30 | 9.34 | 0.05 | | | |
| 50.02 | 11.31 | 0.05 | | | | 121.03 | 9.34 | 0.05 | | | |
| 50.34 | 11.30 | 0.05 | | | | 122.51 | 9.31 | 0.05 | | | |
| 50.43 | 11.27 | 0.05 | | | | 125.30 | 9.27 | 0.05 | | | |
| 50.54 | 11.26 | 0.05 | | | | 128.08 | 9.22 | 0.05 | | | |
| 50.56 | 11.25 | 0.05 | | | | 130.70 | 9.18 | 0.05 | | | |
| 50.62 | 11.25 | 0.05 | | | | 133.22 | 9.15 | 0.05 | | | |
| 51.04 | 11.23 | 0.05 | | | | 135.69 | 9.11 | 0.05 | | | |
| 50.90 | 11.20 | 0.05 | | | | 138.13 | 9.07 | 0.05 | | | |
| 50.63 | 11.20 | 0.05 | | | | 140.45 | 9.05 | 0.05 | | | |
| 51.48 | 11.18 | 0.05 | | | | 146.36 | 8.96 | 0.05 | | | |
| 51.93 | 11.16 | 0.05 | | | | | | | | | |
| 51.62 | 11.15 | 0.05 | | | | | | | | | |
| 52.90 | 11.14 | 0.05 | | | | | | | | | |
| 52.98 | 11.13 | 0.05 | | | | | | | | | |
| 52.89 | 11.11 | 0.05 | | | | | | | | | |
| 53.51 | 11.09 | 0.05 | | | | | | | | | |
| 54.00 | 11.09 | 0.05 | | | | | | | | | |
| 53.96 | 11.08 | 0.05 | | | | | | | | | |
| 53.85 | 11.05 | 0.05 | | | | | | | | | |
| 54.68 | 11.04 | 0.05 | | | | | | | | | |
| 54.98 | 11.03 | 0.05 | | | | | | | | | |
| 54.93 | 11.01 | 0.05 | | | | | | | | | |
| 55.45 | 11.00 | 0.05 | | | | | | | | | |
| 55.24 | 10.97 | 0.05 | | | | | | | | | |
| 55.93 | 10.96 | 0.05 | | | | | | | | | |
| 56.04 | 10.95 | 0.05 | | | | | | | | | |
| 56.59 | 10.95 | 0.05 | | | | | | | | | |
| 56.61 | 10.93 | 0.05 | | | | | | | | | |
| 56.69 | 10.92 | 0.05 | | | | | | | | | |
| 57.09 | 10.91 | 0.05 | | | | | | | | | |
| 57.50 | 10.90 | 0.05 | | | | | | | | | |
| 57.76 | 10.88 | 0.05 | | | | | | | | | |
| 58.40 | 10.88 | 0.05 | | | | | | | | | |
| 58.40 | 10.87 | 0.05 | | | | | | | | | |
| 58.36 | 10.86 | 0.05 | | | | | | | | | |
| 58.91 | 10.85 | 0.05 | | | | | | | | | |
| 59.34 | 10.84 | 0.05 | | | | | | | | | |
| 59.25 | 10.83 | 0.05 | | | | | | | | | |
| 59.78 | 10.81 | 0.05 | | | | | | | | | |
| 59.48 | 10.81 | 0.05 | | | | | | | | | |
| 60.72 | 10.80 | 0.05 | | | | | | | | | |
| 60.52 | 10.79 | 0.05 | | | | | | | | | |
| 60.81 | 10.78 | 0.05 | | | | | | | | | |
| 61.17 | 10.77 | 0.05 | | | | | | | | | |
| 61.04 | 10.75 | 0.05 | | | | | | | | | |
| 61.91 | 10.74 | 0.05 | | | | | | | | | |
| 62.68 | 10.73 | 0.05 | | | | | | | | | |
| 62.28 | 10.72 | 0.05 | | | | | | | | | |
| 62.12 | 10.71 | 0.05 | | | | | | | | | |
| 63.27 | 10.71 | 0.05 | | | | | | | | | |
| 63.59 | 10.70 | 0.05 | | | | | | | | | |
| 62.82 | 10.68 | 0.05 | | | | | | | | | |
| 63.49 | 10.67 | 0.05 | | | | | | | | | |
| 63.80 | 10.67 | 0.05 | | | | | | | | | |
| 63.65 | 10.66 | 0.05 | | | | | | | | | |
| 64.46 | 10.65 | 0.05 | | | | | | | | | |
| 64.50 | 10.64 | 0.05 | | | | | | | | | |
| 63.94 | 10.61 | 0.05 | | | | | | | | | |
| 64.43 | 10.62 | 0.05 | | | | | | | | | |
| 64.80 | 10.59 | 0.05 | | | | | | | | | |
| 64.53 | 10.59 | 0.05 | | | | | | | | | |
| 65.15 | 10.58 | 0.05 | | | | | | | | | |
| 65.93 | 10.57 | 0.05 | | | | | | | | | |
| 65.74 | 10.56 | 0.05 | | | | | | | | | |
| 65.75 | 10.55 | 0.05 | | | | | | | | | |
| 65.50 | 10.54 | 0.05 | | | | | | | | | |
| 65.63 | 10.53 | 0.05 | | | | | | | | | |
| 65.93 | 10.52 | 0.05 | | | | | | | | | |
| 66.01 | 10.52 | 0.05 | | | | | | | | | |
| 66.55 | 10.51 | 0.05 | | | | | | | | | |
| 66.51 | 10.50 | 0.05 | | | | | | | | | |
| 66.98 | 10.49 | 0.05 | | | | | | | | | |
| 67.21 | 10.49 | 0.05 | | | | | | | | | |
| 67.49 | 10.48 | 0.05 | | | | | | | | | |
| 67.87 | 10.47 | 0.05 | | | | | | | | | |
| 68.26 | 10.47 | 0.05 | | | | | | | | | |
| 68.29 | 10.46 | 0.05 | | | | | | | | | |
| 67.84 | 10.45 | 0.05 | | | | | | | | | |
| 68.05 | 10.44 | 0.05 | | | | | | | | | |
| 69.24 | 10.43 | 0.05 | | | | | | | | | |
| 69.93 | 10.43 | 0.05 | | | | | | | | | |
| 69.40 | 10.43 | 0.05 | | | | | | | | | |
| 68.70 | 10.42 | 0.05 | | | | | | | | | |
| 69.19 | 10.41 | 0.05 | | | | | | | | | |
| 69.41 | 10.40 | 0.05 | | | | | | | | | |
| 69.49 | 10.40 | 0.05 | | | | | | | | | |
| 71.75 | 10.39 | 0.05 | | | | | | | | | |
| 69.91 | 10.38 | 0.05 | | | | | | | | | |
| 71.10 | 10.38 | 0.05 | | | | | | | | | |
| 69.77 | 10.38 | 0.05 | | | | | | | | | |
| 71.74 | 10.37 | 0.05 | | | | | | | | | |
| 71.49 | 10.36 | 0.05 | | | | | | | | | |
| 70.45 | 10.35 | 0.05 | | | | | | | | | |
| 73.14 | 10.35 | 0.05 | | | | | | | | | |
| 72.28 | 10.34 | 0.05 | | | | | | | | | |
| 72.00 | 10.33 | 0.05 | | | | | | | | | |
| 72.81 | 10.33 | 0.05 | | | | | | | | | |
| 73.34 | 10.32 | 0.05 | | | | | | | | | |
| 73.49 | 10.32 | 0.05 | | | | | | | | | |
| 72.24 | 10.31 | 0.05 | | | | | | | | | |
| 74.17 | 10.31 | 0.05 | | | | | | | | | |
| 73.63 | 10.30 | 0.05 | | | | | | | | | |
| 73.44 | 10.29 | 0.05 | | | | | | | | | |
| 74.35 | 10.29 | 0.05 | | | | | | | | | |
| 74.81 | 10.29 | 0.05 | | | | | | | | | |
| 74.66 | 10.28 | 0.05 | | | | | | | | | |
| 75.11 | 10.27 | 0.05 | | | | | | | | | |
| 74.65 | 10.27 | 0.05 | | | | | | | | | |
| 74.10 | 10.26 | 0.05 | | | | | | | | | |
| 76.47 | 10.26 | 0.05 | | | | | | | | | |
| 75.14 | 10.26 | 0.05 | | | | | | | | | |
| 74.98 | 10.25 | 0.05 | | | | | | | | | |
| 75.56 | 10.25 | 0.05 | | | | | | | | | |
| 76.66 | 10.24 | 0.05 | | | | | | | | | |
| 76.47 | 10.24 | 0.05 | | | | | | | | | |
| 76.55 | 10.23 | 0.05 | | | | | | | | | |
| 77.47 | 10.22 | 0.05 | | | | | | | | | |
| 76.98 | 10.22 | 0.05 | | | | | | | | | |
| 77.49 | 10.21 | 0.05 | | | | | | | | | |
| 78.24 | 10.21 | 0.05 | | | | | | | | | |
| 77.41 | 10.20 | 0.05 | | | | | | | | | |
| 77.53 | 10.20 | 0.05 | | | | | | | | | |
| 78.09 | 10.19 | 0.05 | | | | | | | | | |
| 78.38 | 10.19 | 0.05 | | | | | | | | | |
| 78.26 | 10.19 | 0.05 | | | | | | | | | |
| 77.82 | 10.18 | 0.05 | | | | | | | | | |
| 78.91 | 10.17 | 0.05 | | | | | | | | | |
| 78.65 | 10.17 | 0.05 | | | | | | | | | |
| 79.04 | 10.17 | 0.05 | | | | | | | | | |
| 78.49 | 10.16 | 0.05 | | | | | | | | | |
| 79.74 | 10.16 | 0.05 | | | | | | | | | |
| 79.34 | 10.16 | 0.05 | | | | | | | | | |
| 80.44 | 10.15 | 0.05 | | | | | | | | | |
| 79.79 | 10.15 | 0.05 | | | | | | | | | |
| 80.38 | 10.14 | 0.05 | | | | | | | | | |
| 79.57 | 10.13 | 0.05 | | | | | | | | | |
| 80.81 | 10.13 | 0.05 | | | | | | | | | |
| 79.66 | 10.12 | 0.05 | | | | | | | | | |
| 81.10 | 10.12 | 0.05 | | | | | | | | | |
| 81.42 | 10.11 | 0.05 | | | | | | | | | |
| 79.89 | 10.11 | 0.05 | | | | | | | | | |
| 80.53 | 10.11 | 0.05 | | | | | | | | | |
| 81.13 | 10.10 | 0.05 | | | | | | | | | |
| 81.06 | 10.10 | 0.05 | | | | | | | | | |
| 80.88 | 10.10 | 0.05 | | | | | | | | | |
| 82.38 | 10.09 | 0.05 | | | | | | | | | |
| 82.66 | 10.09 | 0.05 | | | | | | | | | |
| 81.90 | 10.08 | 0.05 | | | | | | | | | |
| 81.98 | 10.08 | 0.05 | | | | | | | | | |
| 81.46 | 10.07 | 0.05 | | | | | | | | | |
| 82.14 | 10.07 | 0.05 | | | | | | | | | |
| 82.91 | 10.06 | 0.05 | | | | | | | | | |
| 82.38 | 10.06 | 0.05 | | | | | | | | | |
| 82.46 | 10.06 | 0.05 | | | | | | | | | |
| 81.82 | 10.05 | 0.05 | | | | | | | | | |
| 83.53 | 10.05 | 0.05 | | | | | | | | | |
| 82.59 | 10.04 | 0.05 | | | | | | | | | |
| 83.61 | 10.04 | 0.05 | | | | | | | | | |
| 85.02 | 10.04 | 0.05 | | | | | | | | | |
| 83.50 | 10.03 | 0.05 | | | | | | | | | |
| 85.17 | 10.03 | 0.05 | | | | | | | | | |
| 83.59 | 10.02 | 0.05 | | | | | | | | | |
| 83.90 | 10.01 | 0.05 | | | | | | | | | |
| 84.58 | 10.01 | 0.05 | | | | | | | | | |
| 84.03 | 10.01 | 0.05 | | | | | | | | | |
| 85.18 | 10.00 | 0.05 | | | | | | | | | |
| 84.62 | 10.00 | 0.05 | | | | | | | | | |
| 84.75 | 9.99 | 0.05 | | | | | | | | | |
| 85.35 | 9.99 | 0.05 | | | | | | | | | |
| 85.72 | 9.98 | 0.05 | | | | | | | | | |
| 85.86 | 9.98 | 0.05 | | | | | | | | | |
| 85.55 | 9.98 | 0.05 | | | | | | | | | |
| 84.45 | 9.98 | 0.05 | | | | | | | | | |
| 85.71 | 9.97 | 0.05 | | | | | | | | | |
| 86.51 | 9.97 | 0.05 | | | | | | | | | |
| 85.60 | 9.96 | 0.05 | | | | | | | | | |
| 85.75 | 9.95 | 0.05 | | | | | | | | | |
| 87.25 | 9.95 | 0.05 | | | | | | | | | |
| 87.69 | 9.95 | 0.05 | | | | | | | | | |
| 86.99 | 9.95 | 0.05 | | | | | | | | | |
| 87.43 | 9.94 | 0.05 | | | | | | | | | |
| 86.90 | 9.93 | 0.05 | | | | | | | | | |
| 86.28 | 9.93 | 0.05 | | | | | | | | | |
| 88.38 | 9.93 | 0.05 | | | | | | | | | |
| 88.24 | 9.92 | 0.05 | | | | | | | | | |
| 87.76 | 9.92 | 0.05 | | | | | | | | | |
| 87.66 | 9.92 | 0.05 | | | | | | | | | |
| 87.58 | 9.91 | 0.05 | | | | | | | | | |
| 87.60 | 9.91 | 0.05 | | | | | | | | | |
| 88.15 | 9.90 | 0.05 | | | | | | | | | |
| 88.36 | 9.90 | 0.05 | | | | | | | | | |
| 88.30 | 9.90 | 0.05 | | | | | | | | | |
| 89.05 | 9.89 | 0.05 | | | | | | | | | |
| 87.93 | 9.89 | 0.05 | | | | | | | | | |
| 89.78 | 9.89 | 0.05 | | | | | | | | | |
| 88.59 | 9.88 | 0.05 | | | | | | | | | |
| 88.59 | 9.88 | 0.05 | | | | | | | | | |
| 89.53 | 9.88 | 0.05 | | | | | | | | | |
| 89.92 | 9.87 | 0.05 | | | | | | | | | |
| 90.42 | 9.87 | 0.05 | | | | | | | | | |
| 90.06 | 9.86 | 0.05 | | | | | | | | | |
| 90.22 | 9.86 | 0.05 | | | | | | | | | |
| 90.64 | 9.86 | 0.05 | | | | | | | | | |
| 88.79 | 9.85 | 0.05 | | | | | | | | | |
| 90.36 | 9.85 | 0.05 | | | | | | | | | |
| 90.27 | 9.85 | 0.05 | | | | | | | | | |
| 89.55 | 9.84 | 0.05 | | | | | | | | | |
| 89.69 | 9.83 | 0.05 | | | | | | | | | |
| 92.41 | 9.84 | 0.05 | | | | | | | | | |
| 92.42 | 9.83 | 0.05 | | | | | | | | | |
| 90.79 | 9.83 | 0.05 | | | | | | | | | |
| 92.96 | 9.82 | 0.05 | | | | | | | | | |
| 90.70 | 9.81 | 0.05 | | | | | | | | | |
| 92.40 | 9.81 | 0.05 | | | | | | | | | |